\documentclass[namedate,webpdf,contemporary,mediumone]{fua-preprint}

\usepackage{booktabs}   
\usepackage{placeins}   

\usepackage{setspace}

\makeatletter \def\societylogo{} \makeatother

\newif\ifJoEGbuild
\ifJoEGbuild\else
\makeatletter
\def\@journaltitle{}\def\@DOI{}
\def\ps@headings{%
  \def\@evenhead{\hbox to \textwidth{\textbf{\thepage}\hfill}}%
  \def\@oddhead{\hbox to \textwidth{\hfill\textbf{\thepage}}}%
  \def\@evenfoot{}%
  \def\@oddfoot{}}
\def\ps@opening{%
  \def\@evenhead{}\def\@oddhead{}%
  \def\@evenfoot{}\def\@oddfoot{}}
\makeatother
\AtBeginDocument{\pagestyle{headings}}
\fi

\providecommand{\today}{\ifcase\month\or January\or February\or March\or April\or
  May\or June\or July\or August\or September\or October\or November\or December\fi
  \space\number\day, \number\year}

\newcommand{\Oinfo}{\mathcal{O}}

\newcommand{\fuan}{8{,}808}            
\newcommand{\nbeta}{498{,}839}         

\newcommand{\PaperTitleBreak}{ }
\newcommand{\PaperTitle}{Beyond city pairs:\PaperTitleBreak Higher-order information in urban\PaperTitleBreak growth across borders}
\newcommand{\PaperShortTitle}{Beyond city pairs}
\newcommand{\FirstAuthor}{Cillian Hourican}
\newcommand{\SecondAuthor}{Debraj Roy}
\newcommand{\AuthorDepartment}{Computational Science Lab}
\newcommand{\AuthorInstitution}{University of Amsterdam}
\newcommand{\AuthorCountry}{The Netherlands}
\newcommand{\CorrespondingEmail}{c.hourican@uva.nl}

\ifdefined\AnonymousReview
\hypersetup{pdftitle={\PaperTitle},pdfauthor={}}
\else
\hypersetup{pdftitle={\PaperTitle},pdfauthor={\FirstAuthor\space and \SecondAuthor}}
\fi
\begin{document}

\copyrightyear{2026}
\pubyear{2026}
\firstpage{1}

\renewcommand{\PaperTitleBreak}{\\}
\title[\PaperShortTitle]{\PaperTitle}

\ifdefined\AnonymousReview
\author{}
\else
\author[1,$\ast$]{\FirstAuthor}
\author[1]{\SecondAuthor}
\address[1]{\orgdiv{\AuthorDepartment}, \orgname{\AuthorInstitution},
  \orgaddress{\country{\AuthorCountry}}}
\corresp[$\ast$]{Corresponding author.
  \href{mailto:\CorrespondingEmail}{\CorrespondingEmail}}
\fi

\abstract{We apply higher-order information measures to urban growth across borders and
spatial units. We analyse 8,808 functional urban areas using gridded GDP
estimates allocated from national accounts. Negative estimated O-information
occurs in 29.93\% of cross-border triplets and 2.00\% of domestic triplets.
This fifteenfold contrast falls to fourfold at comparable dependence strength
and persists under data-quality restrictions. Target-specific decompositions
show how information overlaps or combines across cities. Administrative
subdivisions have mean O-information within matched random-map ranges
inside countries. These findings show why interpreting geographic dependence
requires specifying comparison cities and spatial units, while economic
connections remain to be established independently.
}



\maketitle

\section{Introduction}
\label{sec:introduction}

Imagine three neighbouring cities whose economies rise and fall together.
They may share a national shock, one city's activity may spill into another,
or the combination of two cities may help explain a third in a way that
neither explains well alone. These possibilities matter for regional
integration and the spread of shocks. A correlation between two cities gives
only part of the description; observing all three lets us ask how much the
information supplied by the two neighbours overlaps, and what their
combination adds. These relationships among three places are the focus of
our description of information shared across regional economies.

National borders make this question especially relevant. Research on trade
shows that proximity alone does not explain the strength of cross-border
connections \citep{mccallum1995,anderson2003}. Yet a weak bilateral growth
relationship need not mean that a foreign city's growth is uninformative:
its contribution may become apparent only alongside another city's series.
We ask how these combinations differ within and across countries, and how
their description changes when the same economic surface is divided into
different spatial units.

We study estimated GDP per capita for 1992--2019 across \fuan\ functional
urban areas (FUAs) worldwide. Throughout the paper, ``city'' means an
estimated FUA: an urban centre and its modelled surrounding commuting area,
rather than a municipality \citep{ghsfua2019,morenomonroy2021}. Annual changes in these
estimates give us 27 growth observations per city. Population is fixed at
2015, so these changes also equal growth in attributed total GDP; they do
not incorporate annual population changes. The GDP estimates come
from a gridded product that allocates national economic activity using
nighttime lights \citep{chen2022gdp}; they provide a common basis for comparing
places. National allocation can itself generate shared city dynamics, so
we study dependence in these estimates without equating it with observed
flows between local economies.

\begin{figure}[p]
\centering
\includegraphics[width=\textwidth,height=.58\textheight,keepaspectratio]{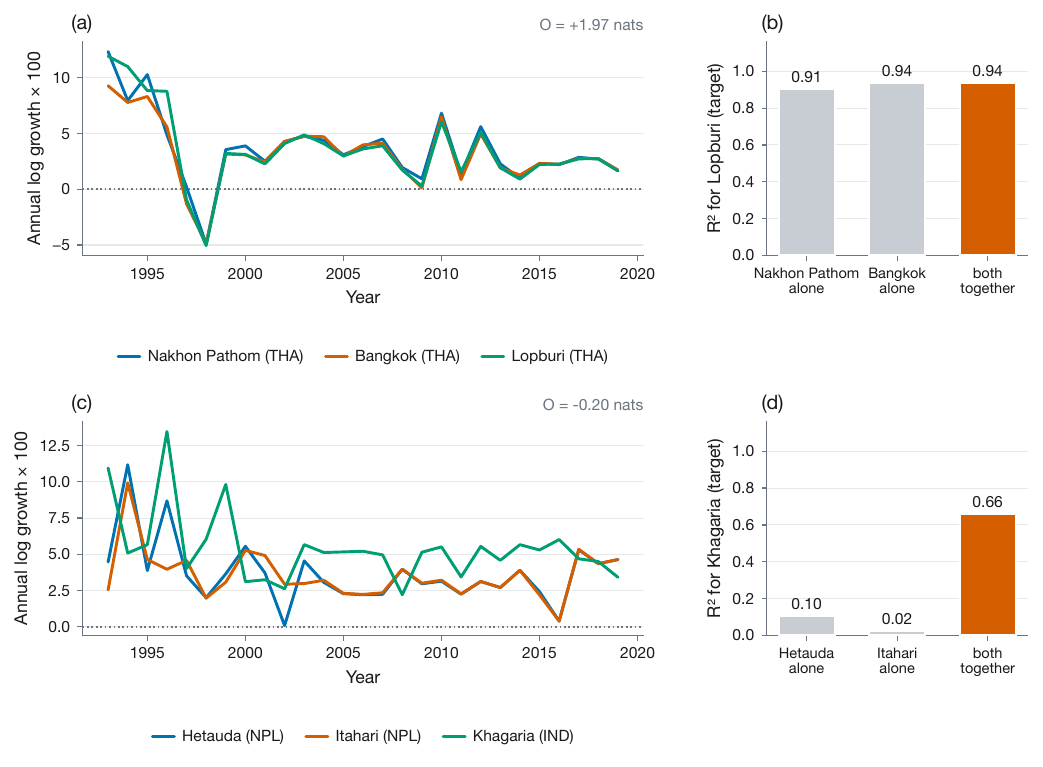}
\caption{Why two places at a time are not enough. First read the growth
  paths on the left, then compare the separate fits on the right, and finally
  the fit using both sources. Top row: Nakhon Pathom and Bangkok each explain
  much of target Lopburi's growth variation; combining them adds little,
  illustrating redundancy. Bottom row: Hetauda and Itahari in Nepal each
  explain little of target Khagaria's growth in India, but together explain
  $66\%$, illustrating synergy. Panels (a,c) show annual GDP-per-capita log
  growth multiplied by 100 for 1993--2019. Panels (b,d) show in-sample $R^2$
  from linear regressions with an intercept and either one source or two
  additive sources; no multiplicative term is used. The opposite-signed
  slopes in the Khagaria joint fit use differences between the source
  paths. O-information annotations indicate
  a redundant balance in the top row and a synergistic balance below.
  Those annotations use the Gaussian-copula estimator; the regressions
  use the untransformed log-growth series. These deliberately selected
  examples illustrate the interpretation, not typical or statistically
  significant individual triplets.}
\label{fig:city-examples}
\end{figure}

Figure~\ref{fig:city-examples} introduces the distinction in three steps.
First, look at the annual growth paths in the top row. Nakhon Pathom,
Bangkok and Lopburi are all in Thailand, and their paths rise and fall
closely together. We take Lopburi as the target: the city whose growth we
want to describe using the other two, which we call sources.
Second, fit Lopburi's growth using each source separately. The bars report
$R^2$, the fraction of its growth variation explained by the fitted
relationship in those same years. Nakhon Pathom alone explains $90.6\%$;
Bangkok alone explains $93.6\%$. Third, include both sources in the same
regression. The fit rises only to $93.7\%$. Knowing the second series adds
little once the first is known, because much of their information overlaps.
This is the intuition behind \emph{redundancy}.

Now repeat these steps for the bottom row. Hetauda and Itahari are in Nepal,
while the target, Khagaria, is in India. Each Nepali city's growth explains
little of Khagaria's variation on its own: $10.5\%$ and $2.3\%$, respectively.
Yet using both raises the explained fraction to $66.0\%$. The combination
is useful even though neither source individually offers a strong fit.
This is the intuition behind \emph{synergy}: information becomes available
through considering the sources jointly.

What changes in the joint regression? We use an intercept and two additive
growth terms,
\begin{equation}
  g_{\mathrm{target},t}=\alpha+\beta_1g_{\mathrm{source\,1},t}
  +\beta_2g_{\mathrm{source\,2},t}+\varepsilon_t.
\end{equation}
There is no product or interaction term. For Khagaria the fitted slopes
are $+1.564$ on Hetauda and $-1.887$ on Itahari. The joint fit can therefore
use a contrast between the Nepali growth paths that neither single-source
fit isolates. A large improvement does not require a nonlinear regression.
These are descriptions of 27 observed years, not tests of prediction in new
years or evidence that one city's activity causes another's.

O-information summarises the balance between redundancy and synergy within
a triplet: positive towards overlapping information, negative towards joint
information. Partial
information decomposition (PID) then separates the information about a
chosen target city into shared, source-specific and joint contributions.
The examples introduce these ideas; Section~\ref{sec:methods} explains how
they are measured and how the cases were selected.

Information measures have been used to describe configurations of regional
innovation \citep{leydesdorff2008,leydesdorff2014} and joint dependencies in
financial markets \citep{scagliarini2020,scagliarini2022}. We apply established
higher-order information measures to the geography of urban growth,
distinguishing overlapping, unique and jointly supplied information.
The application examines two choices that shape descriptions
of national and cross-border connections. First, an isolated bilateral
relationship can understate the information available when a third place
is observed: shared national exposure may overlap across cities, while
their differences help describe a foreign target. Second, a comparison of
maps can attribute a change in information to boundary placement when
national composition and coverage change at the same time. We make these
choices explicit using a familiar covariance-based description of growth,
rather than treating an information score as a direct measure of integration.

The geographic unit is part of the measurement problem. Combining cities into
regions changes their time series and relationships. Drawing different
boundaries may also group places with different national exposures. This is
the modifiable areal unit problem (MAUP): results can depend on both the size
of the units and the way their boundaries are drawn
\citep{openshaw1984,briant2010,dapena2016}.
The question is therefore not only whether a pattern appears among cities,
but whether it survives comparisons that change the map while holding other
features of the measurement reasonably constant.

The city-level results show a substantial descriptive contrast.
Of nearby triplets wholly within one country, $2.00\%$ have negative
estimated O-information. Among triplets
crossing national borders, the proportion is $29.93\%$, approximately
fifteen times as high. Comparing the groups at similar overall dependence
reduces this ratio to approximately four. Each geographically eligible
triplet receives equal weight, so these are proportions of city combinations,
not of countries or economic output. Most triplets overall have positive scores.
This pattern motivates the geographic comparison: we ask how the balance
of growth information varies across borders and spatial units. Explaining
its economic origins requires further evidence about the places being
compared and the construction of their growth estimates.

Two further findings qualify the border contrast. First, the decomposition
locates a foreign city's information relative to a domestic comparison city:
the latter usually supplies the largest unique component under our chosen
rule, while their combination can add further information. Second,
administrative subdivisions have mean O-information
within the range of comparable random subdivisions drawn inside the same
countries, across three regional scales. The national border contrast and
the subdivision result answer different questions: the former compares city
configurations across countries, while the latter varies regional boundaries
with countries held fixed. Together they show why dependence strength,
information shared among cities, and the choice of map need to be considered
separately.

The paper proceeds from economic interpretation and methods to city-level
patterns, their decomposition, and the partition comparison. We then examine
associations with economic outcomes and discuss the implications and limits.

\FloatBarrier

\section{Reading growth dependence in economic geography}
\label{sec:framework}

Regional economies are connected through production, knowledge, institutions
and shared exposure to shocks. Relational and evolutionary economic geography
asks how these connections shape development
\citep{bathelt2003,gluckler2007,frenken2007}.
Related variety, for example, concerns the opportunities for learning among
activities that share some capabilities while differing in others
\citep{frenken2007}. These connections motivate our question but are not
measured by the growth panel. An information decomposition asks a narrower empirical
question: do the observed growth series repeat similar information or become
more informative in combination? Dependence between growth series can reflect many such processes,
rather than a direct measure of any one. Our contribution is to describe how
information overlaps across places and examine how that description depends
on the comparison city and spatial units.

\emph{Redundancy} means that two source cities carry overlapping information
about a target city. A common national business cycle is one possible source.
The Thai example illustrates this: either neighbouring series already
captures most of the variation that the two capture jointly. Redundancy can
be consistent with shared exposure, but it does not establish whether the
shared driver is policy, trade, industrial composition or the construction
of the GDP data.

\emph{Unique information} is information about the target available from one
source alone and not shared with the other. Its size depends on which comparison
city we choose. If a target's domestic neighbour carries more unique
information, it offers a different description of the target from the
foreign source; the latter can still add information through synergy. This is descriptive accounting, not a
demonstration of a directed economic spillover. Source and target are roles
in a decomposition of contemporaneous series, not evidence of a causal arrow.
The extra information obtained by adding a source after observing the other
includes both its unique contribution and synergy.

\emph{Synergy} is information supplied by the two sources together beyond
their redundant and unique contributions. The Nepal--India example shows
how subtracting a shared component from two growth paths can reveal a
relationship with a third place. This happens within an ordinary linear
regression; it does not require complementary production or establish a
cross-border economic connection.

The Triple-Helix literature measures configurations of such dimensions as
geography, technology and organisation \citep{leydesdorff2008,leydesdorff2014}.
Here the variables are instead growth series from three distinct places.
Financial applications also study joint information, including its temporal
transfer \citep{scagliarini2020,scagliarini2022}; our main comparisons describe
contemporaneous growth. This complements studies of growth and
convergence using the same FUA data
\citep{mengesha2026meso,mengesha2026regimes}, and relates to evidence that
urban and regional findings depend on their areal definition
\citep{arcaute2015,dapena2016}.
We do not claim that a Gaussian information measure uncovers dependence
absent from the full covariance matrix: its contribution is the organisation
and interpretation of those dependencies across places.

Changing the map changes the object being described. Aggregating raster GDP
over a region changes the growth series; averaging constituent cities'
growth is a different construction. We distinguish these exercises below.
A boundary separating countries may also separate national drivers. Pairwise and three-variable
summaries can both change. Whether one is more sensitive than the other is
an empirical question, requiring the same transform, geographic support and
aggregation rules on both sides. Neither a small O-information value nor its
decline under aggregation establishes that dependence has disappeared.

\FloatBarrier

\section{Data and methods}
\label{sec:methods}

The analysis proceeds from cities to relationships and then to maps. We
assemble annual city growth series, describe nearby triplets, compare
pairs across borders, decompose the information supplied by alternative
comparison cities, and repeat the measurement under different spatial
partitions. This section explains what is measured and how to read the
figures. Estimator formulas and detailed sample rules are in the supplement.

\subsection{Which places and years are compared?}
\label{subsec:data}

We use the 2015 boundaries of the GHSL estimated functional urban areas
\citep{ghsfua2019}.
Their surrounding commuting areas are modelled from geographic and
population characteristics; commuting flows are not observed worldwide.
The attributed panel used by \citet{mengesha2026regimes} contains 8,809 FUAs.
Requiring strictly positive GDP per capita in every year from 1992 to 2019
leaves \fuan. Annual growth is
\begin{equation}
 g_{it}=\log y_{it}-\log y_{i,t-1},
\end{equation}
where $y_{it}$ divides attributed annual GDP by the city's fixed 2015
GHS-POP population. The 28 levels provide 27 annual growth observations
for 1993--2019. With a fixed denominator, these changes also equal log
changes in attributed total GDP; they do not incorporate annual population
change. The population product and aggregation are specified in
Supplement~\ref{app:sample-construction}.

The GDP surface allocates national economic activity using calibrated
nighttime lights \citep{chen2022gdp}. It is not an independently observed
city account in every location. Some countries have very little spatial
variation in their attributed city growth. We therefore report both the
full reference sample and a restricted sample that excludes countries with
almost indistinguishable city growth paths and relationships with nearly
singular correlation matrices. The two rules address different problems:
limited spatial detail in the data and numerical instability in measuring
dependence. Neither selects observations by the sign of O-information.
The country rule excludes 872 cities in 18 country or territory codes;
Supplement~\ref{app:dispersion} defines the thresholds; Supplement~\ref{app:corrected-results} reports the sensitivity checks.

Figures~\ref{fig:border-contrast} and~\ref{fig:beta-atoms} retain the full
triplet and PID distributions to describe the complete input geography.
We report the restricted estimates alongside the headline contrast and in
the supplement to establish how much the low-dispersion cases contribute;
agreement does not validate the remaining local GDP estimates. The distance and border maps, matched partition comparisons
and common-city grouping maps use their stated restrictions. These
analyses answer different geographic questions and need not have the same
sample. Table~\ref{tab:sample-rules} summarises their coverage.

\subsection{What does dependence mean here?}
\label{subsec:methods-hoi}

Mutual information measures how much knowing one growth series reduces
uncertainty about another. We call this \emph{growth dependence}, since
both positive and negative associations can be informative: cities need
not grow and slow together. Information quantities use natural logarithms
and are expressed in nats. Zero population mutual information means
independence; greater values indicate stronger dependence.

For a triplet, O-information summarises the balance between redundant and
synergistic information \citep{rosas2019}. A positive value indicates more
overlapping information; a negative value indicates more information
available through combinations. It is not a measure of the total amount
of dependence. Near-zero O-information can accompany weak dependence,
a balance of redundancy and synergy, or dependence concentrated in one
pair. We therefore also report total correlation (TC), which measures
overall dependence, and dual total correlation (DTC), the second dependence
summary in the identity $\Oinfo=TC-DTC$. TC compares joint uncertainty with
the sum of the cities' separate uncertainties. DTC compares joint uncertainty
with the sum of uncertainties left in each city after observing the other two.
Both can be positive even when their difference is zero. They are not
separate PID redundancy and synergy components.

We replace each city's growth observations by normal scores derived from
their ranks, then calculate Gaussian information measures from the
covariance matrix \citep{ince2017}. This reduces sensitivity to marginal
distribution shape, while assuming a Gaussian copula for dependence.
Under this estimator, information is fully determined by correlations
among the transformed series. ``Higher-order'' refers here to the way
information is organised among three cities, rather than dependence
missing from their complete covariance matrix.

O-information, TC, DTC and pairwise mutual information use a finite-sample
entropy correction. This can make estimated mutual information slightly
negative; it does not indicate an inverse association. PID uses plug-in
Gaussian information without that correction. Thus reported negative-score
fractions refer to corrected O-information, not an exactly equivalent count
of plug-in PID balances; the conventions can differ close to zero.
The opening regressions
instead use the original log-growth observations: their $R^2$ values are
in-sample fits from additive linear models, not information amounts.
Supplement~\ref{app:estimation} specifies these conventions. The two
examples were deliberately selected to illustrate contrasting configurations;
their identities and coefficients are in Supplement~\ref{app:examples}.

\subsection{Which nearby relationships are included?}
\label{subsec:methods-pruning}

The triplet description includes every combination whose three pairwise
separations are at most 1,000~km, giving 428,568,795 triplets. All three
cities must share a country code for the triplet to count as within-country;
otherwise it is cross-border. The distance limit defines a regional
neighbourhood for comparison, rather than an estimated range of interaction.

For pairwise dependence, we compare within-country and cross-country
means in 50~km distance bands. Border maps focus on cities at most 300~km
apart in countries sharing a land boundary. The shared border must be no
farther from their midpoint than the cities are from each other. We retain
borders with at least 30 qualifying pairs and enough domestic information
to calculate the adjustment. This gives the same 87 borders and 19,809
pairs in the raw and adjusted maps. Distance-point conventions and exact
city eligibility rules are in Supplement~\ref{app:sample-construction};
border definitions and alternative references are in Supplement~\ref{app:border-sensitivity}.

\subsection{What does a second city add?}
\label{subsec:methods-pid}
\label{subsec:methods-beta}

Partial information decomposition (PID) chooses a target, the city whose
growth is described, and two sources \citep{williams2010,bertschinger2014}.
It separates information shared by both sources, information unique to
each, and information available only from their combination. In the
systematic comparison, the \emph{source} is a city across the border from
the target and the \emph{context} is a second comparison city. These are
computational roles, not directions of economic influence. Source-unique
information is available from the foreign source alone but not the context
alone; context-unique information is the reverse. Adding a source after
observing the other provides its unique information \emph{plus} synergy.

Each cross-country pair within 500~km is compared against up to five
contexts: a domestic neighbour of either endpoint, a capital-region
representative in either country, or a nearby third-country city. This
uses one fixed target orientation per unordered pair, determined by input
row order, and produces \nbeta\ source--target--context combinations.
The frequencies describe this orientation rather than both directions.
The restricted
sample filters these same combinations, retaining 406,759; it does not
select replacement neighbours or capitals. Detailed selection rules are
in Supplement~\ref{app:sample-construction}.

For these scalar growth variables, the Gaussian decomposition reduces to
the minimum-mutual-information rule: redundancy equals the smaller of the
two source--target information amounts \citep{barrett2015,venkatesh2022}.
The other components follow from the information identities. The weaker
source consequently has zero unique information within each triplet,
although it can contribute through synergy. The rule and estimator conventions
are in Supplement~\ref{app:estimation}.

Figure~\ref{fig:beta-atoms}(a,b) reports information amounts in nats.
Panels (c,d) instead count the combinations in which each component is
largest. Winning in half the combinations does not mean supplying half
their information. We retain this distinction throughout the analysis.

\subsection{How are alternative maps compared?}
\label{subsec:methods-partitions}

We aggregate the underlying GDP and population rasters into administrative
or random regions, then form regional growth series. Population is fixed
at its 2015 value, so it cancels when taking changes in log GDP per capita.
Triplets of regions use the same 27 growth observations, estimator and
1,000~km limit on all three separations as the city comparison.

The matched comparison uses first-level administrative regions worldwide
(ADM1) and two European statistical scales, NUTS-1 and NUTS-2. For each,
we draw 200 random Voronoi subdivisions: each location belongs to its
nearest seed, with cells confined to the same country. Region counts by
country and GDP/population coverage are fixed; spacing and triplet
composition must satisfy the declared matching tolerances. Both NUTS
scales use the same source coverage, including its overseas pieces.
Supplement~\ref{app:maup-design} documents the design.

Within each map, eligible triplets receive equal weight. Each random map
then contributes one mean to the reference distribution, regardless of
its number of triplets. Bands and whiskers span the central 95\% of those
200 map means. They describe variation across alternative maps, rather
than sampling uncertainty in the observed growth series. This comparison
varies subdivisions \emph{within} fixed countries. A separate illustration
compares whole countries with one random map that can cross national
boundaries; its coverage and region counts differ
(Figure~\ref{fig:partition-illustration}).

\subsection{Statistical scope}
\label{subsec:methods-stats}

The information measures are descriptive. We assessed reference systems
with matched total information using the NuMIT approach \citep{liardi2025},
which holds dependence strength fixed while varying its arrangement.
With 27 observations, the resulting tail probabilities are sensitive to
estimating the matching quantity and are not adequately calibrated across
all tested settings. The simulations also have limited resolution for
large families of overlapping triplets. We therefore make no claims of
statistically exceptional individual triplets or PID components. A negative
O-information score describes a balance towards synergy; it is not a
significance decision. Supplement~\ref{app:numit} reports the calibration
and resolution limits supporting this scope; detailed diagnostics are archived.

\FloatBarrier

\section{How growth dependence differs across national borders}
\label{sec:alpha-descriptive}

\subsection{From the global distribution to the border contrast}
\label{subsec:alpha-global-dist}
\label{subsec:alpha-border-contrast}

\begin{figure}[tbp]
\centering
\includegraphics[width=\textwidth,height=.58\textheight,keepaspectratio]{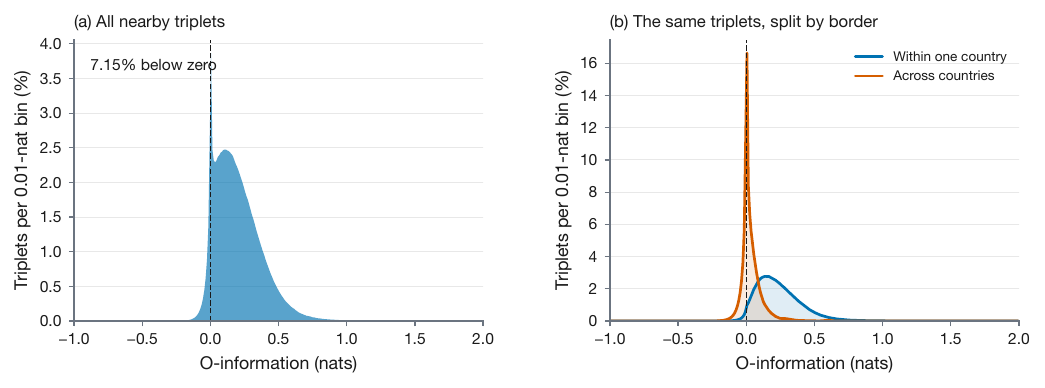}
\caption{The distribution of higher-order growth information, and how it
  differs across national borders. Panel (a) shows all 428,568,795 nearby
  city triplets; panel (b) splits the same population into 349,632,590
  within-country and 78,936,205 cross-border triplets. Gaussian-copula O-information uses
  annual GDP-per-capita log growth for 1993--2019, with all three city
  separations at most 1,000~km. Positive estimated O-information indicates a redundant balance;
  negative values indicate a synergistic balance. The finite-sample
  convention is specified in Section~\ref{subsec:methods-hoi}. Curves join percentages
  in 0.01-nat bins, with the full population or respective full group as
  denominator. The vertical scales differ; values outside the displayed
  $-1$ to $2$ nat range remain in the denominators. Negative scores account
  for $7.15\%$ overall, $2.00\%$ within countries and $29.93\%$ across
  countries. The latter two fractions differ by a factor of $14.94$.
  These are full-reference descriptive proportions before the dispersion
  screen, not significance rates. Screened comparisons are reported in
  Table~\ref{tab:corrected-alpha}.}
\label{fig:border-contrast}
\end{figure}

The opening examples show what overlapping and joint information can look
like. Figure~\ref{fig:border-contrast} now puts them in the population of
nearby city triplets. Panel (a) pools the full reference of 428,568,795
geographically eligible triplets, before the data-quality restrictions.
Mean O-information is $0.206$ nats and the median is $0.172$.
Negative scores occur in $7.15\%$ of triplets; most therefore lie on the
redundant side of the balance. Within-country combinations make up
$81.58\%$ of triplets, so their pattern strongly influences this global
distribution.

Panel (b) separates triplets wholly within one country from those crossing
at least one national border. The 349,632,590 within-country triplets have
mean O-information $0.245$ nats and negative scores in $2.00\%$ of cases.
The 78,936,205 cross-border triplets have mean $0.034$ nats and negative
scores in $29.93\%$. The ratio of the negative-score fractions is $14.94$.
This is a comparison of their signs, not the fraction of statistically
significant triplets. Giving both groups the same distribution of total
correlation, so that their overall dependence is comparable, reduces the
ratio to approximately $4.00$
(Supplement~\ref{app:corrected-results}). Matching dependence strength
therefore attenuates the contrast but does not remove it; country and
spatial differences remain possible explanations.

The data-quality and numerical restrictions retain 408,163,633 triplets. Negative
scores occur in $2.01\%$ of within-country and $27.31\%$ of cross-border
triplets, a ratio of $13.58$; standardising to the same TC bins gives $3.62$.
The direction therefore survives this restriction, although the restriction
changes the represented countries and relationships. It does not identify
a causal border effect (Table~\ref{tab:corrected-alpha}).

Supplement~\ref{app:gdp-participation} also reports GDP-weighted
participation in negative-score triplets. This describes configurations
involving the covered cities; it does not divide world GDP into output
produced by redundancy and synergy.

\subsection{The border gap after accounting for domestic dependence}
\label{subsec:alpha-borders}

The pairwise comparison asks a related question: how informative is one
city's growth about another's? We use Gaussian-copula mutual information,
which measures dependence of either sign. We therefore call it growth
dependence rather than equating it with same-direction co-movement.
In the restricted distance plot, within-country mean mutual information falls
from $0.614$ nats in the nearest 50~km band to $0.355$ in the band centred
at 975~km. Cross-country band means are lower, ranging from $0.069$ to
$0.093$ nats (Figure~\ref{fig:distance-decay}). The comparison holds distance
bands comparable, but country composition differs between the groups.

For an individual cross-border pair, define the raw gap as the pooled
within-country mean mutual information in its 50~km distance band minus
the pair's own mutual information. Positive gaps indicate less dependence
than this domestic reference. Averaging qualifying pairs gives a gap for
each land border. The raw and adjusted maps use the same screened set
of 87 borders and 19,809 pairs (Figure~\ref{fig:border-map-raw}). The
domestic distance reference is also recomputed after screening. However,
the two countries beside a border may themselves have unusually high
or low domestic dependence.

\begin{figure}[p]
\centering
\includegraphics[width=\textwidth,height=.58\textheight,keepaspectratio]{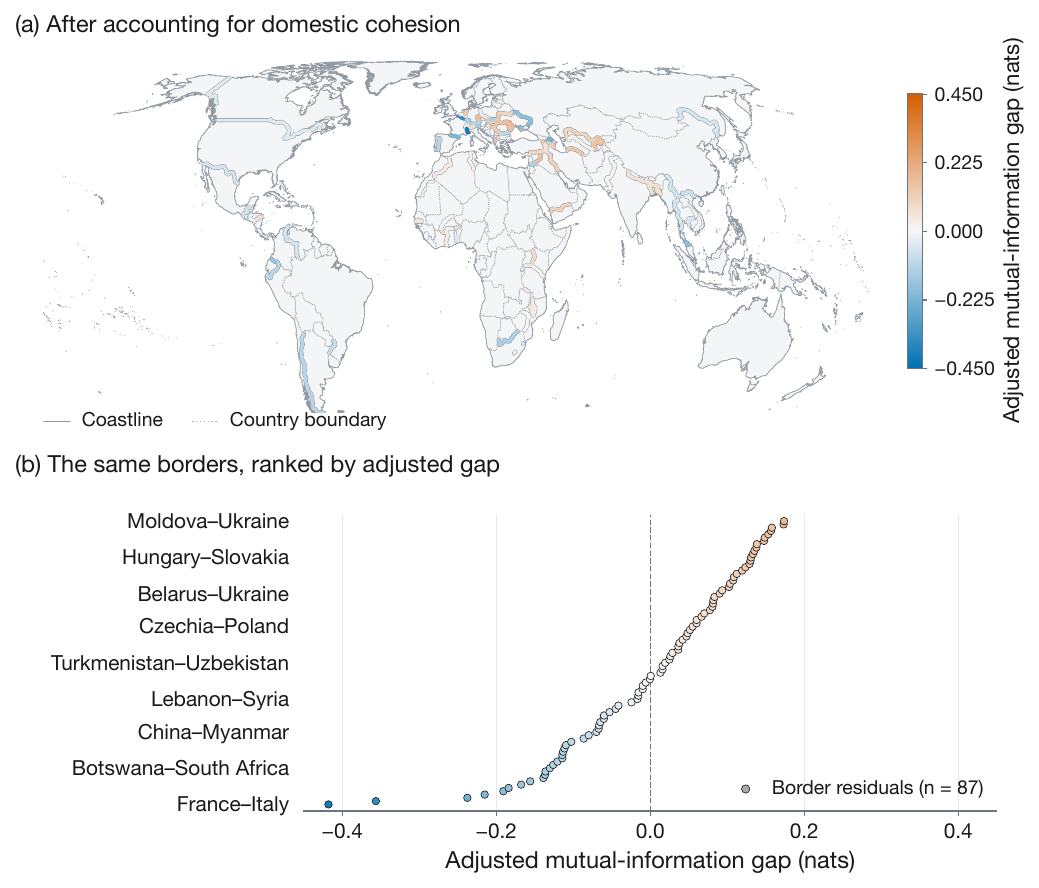}
\caption{The border gap after accounting for each country's domestic
  dependence. Start with each border's mean distance-matched mutual-information
  gap: pooled domestic dependence minus cross-border dependence. Regress
  these means on an intercept and the lower and higher domestic-cohesion
  values of the two adjoining countries. Panels (a,b) map and rank the
  residuals for 87 screened land borders, covering 19,809 city pairs. A positive
  residual means a larger gap than the regression predicts; a negative
  residual means a smaller gap, not necessarily a reversal of the raw gap.
  All information measures use annual GDP-per-capita log growth for
  1993--2019 and are in nats. Pairs are at most 300~km apart, pass the
  border-proximity check, and contribute to borders with at least 30 pairs.
  Country cohesion uses within-country pairs up to 1,000~km. The regression
  weights borders equally. Grey coastlines and dotted boundaries provide
  context; thin outlines identify eligible borders even near zero.
  The map and regional enlargements share a $-0.45$ to $+0.45$
  nat scale without clipping. Dotted-only borders are outside
  the sample. The 300~km
  pooled-reference ranking is descriptive; distance and domestic-reference
  sensitivities are reported in Supplement~\ref{app:border-sensitivity}.}
\label{fig:border-map}
\end{figure}

Figure~\ref{fig:border-map} accounts for this variation descriptively.
We measure each country's domestic cohesion by its mean within-country
mutual information, then fit the raw border gaps against the lower and
higher of the two adjoining countries' cohesion values, with an intercept.
The map shows what remains after subtracting the fitted gap. It includes
87 borders and 19,809 pairs; Germany--Luxembourg is excluded because
Luxembourg has no eligible internal city pair from which to calculate
cohesion. The regression, refitted after screening, explains $33.2\%$ of variation
in raw border gaps.

A positive adjusted gap means cross-border dependence is weaker than this
fitted comparison predicts. A negative adjusted gap means the raw gap is
smaller than predicted. It does not necessarily mean cross-border dependence
exceeds the domestic reference: a positive raw gap can still have a negative
residual. For example, Argentina--Brazil has a positive raw gap of $0.263$ nats
but an adjusted gap of $-0.086$: its cross-border pairs are less dependent
than the pooled domestic reference, yet the gap is smaller than expected
from the two countries' own cohesion. Albania--North Macedonia instead
has a raw gap of $0.485$ and an adjusted gap of $+0.133$ nats. The map
therefore distinguishes a border's raw contrast from its position relative
to domestic dependence. Regional enlargements use the same adjustment and colour scale
(Figure~\ref{fig:border-map-regions}).

The positive raw gap persists under 200 and 400~km distance limits.
Adjusted border rankings are also stable across those limits on common
borders. Using domestic reference pairs from the adjoining countries
instead of the pooled global reference preserves positive mean gaps where
both sides have enough observations, but changes rankings and magnitudes.
The map therefore describes a stated comparison, rather than a unique
ordering of border barriers (Supplement~\ref{app:border-sensitivity}).

Domestic-cohesion adjustment does not control income, industrial structure,
common national shocks or the quality of the spatial GDP estimates. Cities
with nearly identical attributed growth paths can create high domestic
dependence by construction. The map remains a descriptive comparison, and
the pairwise gap and the triplet balance summarise different aspects of
the growth relationships.

\FloatBarrier

\section{What information does the economic context add?}
\label{sec:beta-systematic-pid}

The triplet landscape treats all three cities symmetrically. PID instead
chooses a target and separates what two source cities tell us about its
GDP-per-capita growth. We first return to the opening examples, then apply
the same accounting to cross-border pairs with different context cities.

\subsection{Decomposing the two worked examples}

Take Lopburi as the target and Nakhon Pathom and Bangkok as its sources
(Figure~\ref{fig:beta-atoms}a). Redundancy is the information about Lopburi
available from either source: $1.972$ nats. Bangkok also supplies $0.274$
nats of unique information, while Nakhon Pathom's unique component is zero
at the reported precision. Only $0.002$ nats are supplied through their
combination beyond those shared and unique contributions. This is the
redundancy-dominated example: most of the information overlaps.

For Khagaria, with sources Hetauda and Itahari, the balance is different
(Figure~\ref{fig:beta-atoms}b). The shared component is approximately $0.001$
nats, Itahari's unique component is $0.069$, and Hetauda's is zero.
Their combination supplies another $0.205$ nats through synergy, the largest
component. Hetauda's zero unique component does not make it irrelevant:
once Itahari is known, adding Hetauda also makes the synergistic information
available. The additional information from a source is its unique component
\emph{plus} synergy.

These decompositions use the rank-transformed growth series and are measured
in nats; they do not divide the regression $R^2$ into four parts. The rank
transformation can change the relative strength of the individual source
relationships, as it does for Hetauda and Itahari. The regression examples
and PID therefore illustrate the same broad distinction using different
summaries. Estimator conventions are recorded in
Supplement~\ref{app:examples}.

\subsection{The context changes which information component is largest}
\label{subsec:beta-distributions}

For the systematic comparison, one source is a foreign city and the second
is a comparison city: for example, a domestic neighbour of the target.
We ask which component is largest under each choice of comparison city,
using the fixed pair orientations described in the Methods. The full systematic design contains \nbeta\ decompositions before the additional data-quality restrictions.
Context-unique information is the largest component in $50.75\%$ of decompositions,
redundancy in $26.29\%$, source-unique information in $12.69\%$ and synergy
in $10.26\%$. Their respective means are $0.384$, $0.077$, $0.014$ and
$0.024$ nats. The percentages count which component is largest in each decomposition;
they do not measure its share of the information or how often it is nonzero.

\begin{figure}[htbp]
\centering
\includegraphics[width=\textwidth,height=.58\textheight,keepaspectratio]{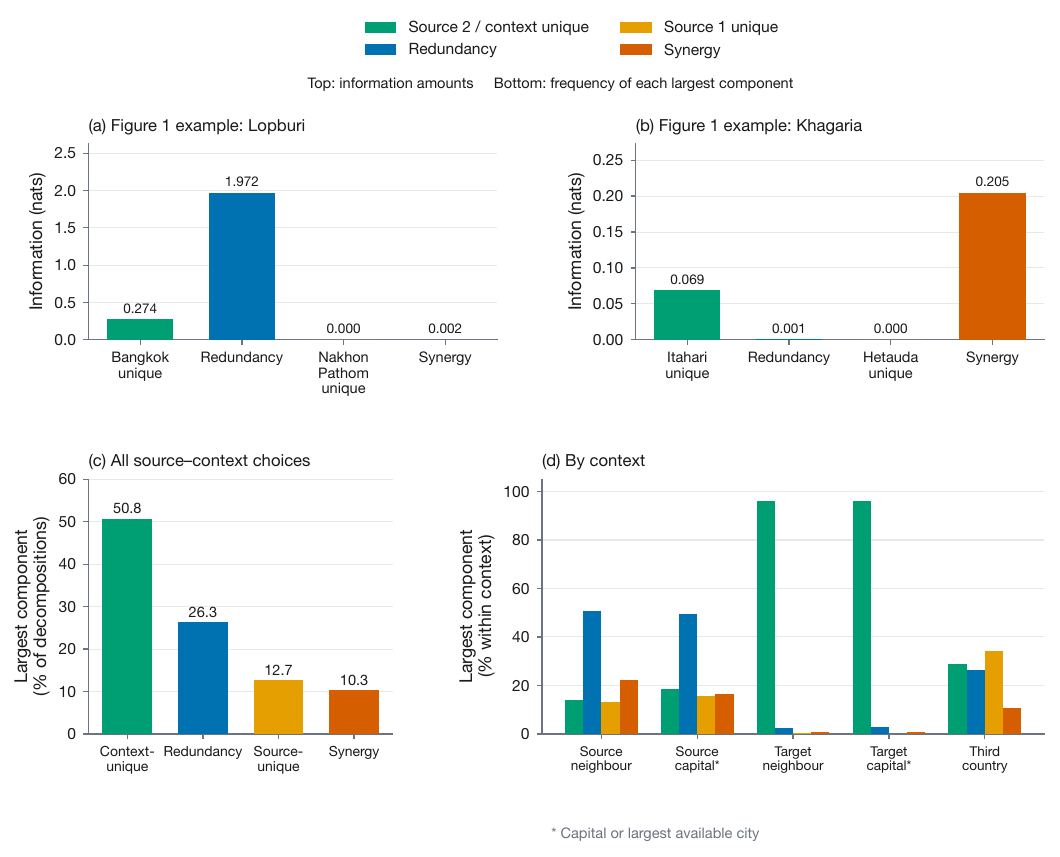}
\caption{\textbf{Decomposing the information shared between cities.}
Panels (a,b) revisit the two triplets in Figure~\ref{fig:city-examples}.
Panel (a) describes Lopburi's growth using Nakhon Pathom (source 1) and
Bangkok (source 2); redundancy is largest. Panel (b) describes Khagaria's
growth using Hetauda (source 1) and Itahari (source 2); synergy is largest.
PID separates information about the target into information shared by both
sources (redundancy), information available only from their combination
(synergy), and information unique to each source. These panels show amounts
in nats, with different vertical scales.
For the systematic comparison in (c,d), each unordered pair has one fixed
target orientation. The \emph{target} is the city whose growth is described, the \emph{source} is a city across the border, and the
\emph{context} is a second comparison city. Source-unique information is
available from the foreign source alone but not from the comparison city
alone; context-unique information is the reverse.
Bars show the percentage of source--target--context combinations in which
each component is largest. Panel (c) pools the comparisons; panel (d)
separates contexts selected as a domestic neighbour of either endpoint,
a capital-region representative in either country, or a third-country city.
Where the capital could not be matched, the country's most populous FUA
in the dataset was used. All panels use Gaussian PID on rank-transformed
annual GDP-per-capita log growth for 1993--2019. Panels (c,d) summarise all
498,839 combinations before the additional data-quality exclusions;
the corresponding restricted results are reported in
Table~\ref{tab:corrected-beta}.}
\label{fig:beta-atoms}
\end{figure}

The context split is more informative than the pooled average
(Figure~\ref{fig:beta-atoms}c,d). When the context is the target's domestic
neighbour or capital context, context-unique information dominates in
$96.2\%$ and $96.1\%$ of decompositions. For the source's domestic neighbour or capital,
redundancy is the most frequent dominant component ($50.7\%$ and $49.3\%$).
The third-country context has a more mixed distribution, with source-unique
information the largest category at $34.1\%$. Under the scalar minimum-information rule, the weaker source has zero
unique information. The target-side domestic result therefore partly
reflects stronger domestic pairwise dependence; it is not independent
evidence of a domestic economic mechanism.
Quantifying the foreign source's additional contribution requires both its
unique information and synergy: together these measure what it adds once
the context is known.

The restricted comparison retains 406,759 decompositions with the original
context choices. Mean synergy is $0.02383$ nats, compared with $0.02396$
in the full reference; it is the largest component in $9.33\%$, compared
with $10.26\%$. Context-unique information remains the most frequent
dominant component ($50.17\%$). Both worked examples pass the declared
restrictions (Supplement~\ref{app:dispersion}).
This pattern is compatible with national exposure and data construction,
as well as with economic relationships; the decomposition cannot separate
those explanations on its own.

\FloatBarrier

\section{How much does the choice of map matter?}
\label{sec:maup}

The city results describe relationships on a fixed map of functional urban
areas. Regional analyses often aggregate economic activity over administrative
units. Here we sum raster GDP within each region before calculating growth;
the supplementary city-grouping illustration instead averages city growth. Does the resulting information pattern reflect a distinctive
administrative geography, or could comparable groupings produce it? This
is a version of the modifiable areal unit problem: both the scale and
boundaries of spatial units can affect measured relationships
\citep{openshaw1984,briant2010,dapena2016}.

\subsection{Administrative subdivisions and comparable random maps}

\begin{figure}[htbp]
\centering
\includegraphics[width=\textwidth,height=.58\textheight,keepaspectratio]{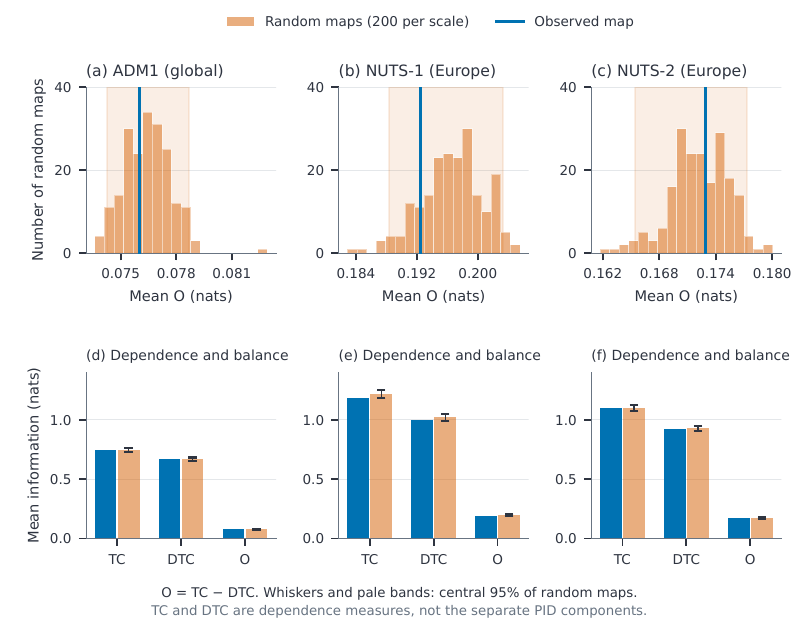}
\caption{\textbf{O-information under administrative and matched random subdivisions.}
Panels (a--c) compare the observed administrative map (blue line) with
200 random partitions at each scale (orange histograms). Each histogram
contains the mean O-information from each random partition. The pale band
spans the central 95\% of random-map values; the observed value lies inside
that band at all three scales. Random boundaries are drawn within countries,
preserving region counts by country and geographic coverage.
Panels (d--f), in the same scale order, show total correlation (TC), dual
total correlation (DTC), and their difference, $O=TC-DTC$. Blue bars show
the observed map and orange bars the average across random maps. Whiskers
span the central 95\% of random-map values. TC summarises overall dependence;
TC and DTC are not separate PID redundancy and synergy components.
All measures are in nats, estimated from regional annual GDP-per-capita
log growth for 1993--2019 (27 observations). Regional GDP per capita is
calculated from the underlying GDP and population rasters, using fixed
2015 population. Triplets have all three separations at most 1,000~km.
After the data-quality and numerical exclusions described in the Methods,
the observed maps retain 2,654 ADM1 regions, 124 NUTS-1 regions and
327 NUTS-2 regions. Both NUTS comparisons use the same European coverage.}
\label{fig:partition-comparison}
\end{figure}

Figure~\ref{fig:partition-comparison} compares observed subdivisions with
200 random maps at each of three scales. All random cells stay inside
national borders and preserve each country's region count and raster
coverage. Thus an ADM1 map is compared with alternative subdivisions of
the same countries, while the two NUTS scales use matched European support.
This controls features that a comparison based only on typical region
spacing would leave free to change.

Read panels (a--c) first. Each histogram shows the mean O-information
obtained from a different random map; the blue line marks the observed
administrative map. At all three scales the observed mean lies inside
the central 95\% of random-map values. For ADM1, observed O-information is
$0.0760$ nats, compared with a random mean of $0.0764$ and a central range
of $0.0743$--$0.0787$. NUTS-1 gives $0.1925$ versus $0.1967$, and NUTS-2
$0.1729$ versus $0.1723$. Observed means are respectively $0.34$ and $1.03$
random-map standard deviations below the random mean, and $0.21$ above it.
The comparisons give no evidence of a large, distinctive reduction in the
redundant balance under administrative subdivisions.

Now read panels (d--f). The low O-information values do not mean little
overall dependence: O-information is the difference between TC and DTC.
For ADM1, observed TC is $0.7455$ nats and DTC is $0.6695$, close to the
random means of $0.7462$ and $0.6698$. NUTS-1 has somewhat lower TC and
DTC than its random maps; both move together, leaving their difference
within the reference range. These component results concern summaries of
overall dependence, not separate PID redundancy and synergy amounts.
Supplement~\ref{app:maup-design} reports the component comparisons,
including their unadjusted tail positions.

The O-information result also holds after standardising every random map
to the observed proportions of triplets spanning one, two or three
countries. Across three scales, six sample restrictions and two weightings,
all 36 observed means remain inside the corresponding central 95\% ranges.
These comparisons condition on the chosen family of random maps. Because
they hold international boundaries fixed, they cannot estimate the effect
of those boundaries or explain how national borders were formed.

\subsection{Contrasts that change the national grouping}

The supplementary maps illustrate why the reference comparison matters.
Figure~\ref{fig:partition-illustration} compares 179 whole countries with
502 regions from one random Voronoi map. Mean O-information is $0.050$
nats for countries and $0.591$ for the random regions. This is a substantial
descriptive contrast between two unmatched maps: coverage, region counts
and national composition all differ. In particular, a random-region triplet
can combine several pieces of one national economy, whereas a triplet of
countries cannot. The contrast reveals how different the measured objects
can be; it does not isolate an effect of boundary placement. By comparison,
Figure~\ref{fig:partition-comparison} holds countries fixed and finds
ordinary map-wide mean O-information within its matched random-map family.
Neither comparison identifies which factor produces the whole-country
contrast.

Figures~\ref{fig:spatial-partitions}
and~\ref{fig:spatial-partitions-minimum} group the same 641 European
cities six ways. They show the mean and minimum triplet score involving
each city or region. Most means are positive, although many places appear
in a negative-score triplet. This illustrates how grouping and averaging
can conceal local combinations. It uses equal-city mean growth and a
300~km neighbourhood, distinct from the raster aggregation and matched
maps in Figure~\ref{fig:partition-comparison}.

\FloatBarrier

\section{Associations with economic outcomes}
\label{sec:delta-mechanism}

Do these information patterns help describe growth outcomes? We relate
2010--2019 log GDP-per-capita growth to initial income, economic composition,
education, governance and each city's fraction of negative-O-information
triplets. Available regional controls leave 465 cities in 25 EU countries.
The participation coefficient is near zero and imprecise: $-0.0011$ on a
zero-to-one scale, with a country-clustered standard error of $0.1336$.
It adds about $0.000001$ to in-sample $R^2$. The outcome overlaps the
information-estimation period, so this is an association rather than a
forecasting exercise (Supplement~\ref{app:delta-convergence-table}).

Supplement~\ref{app:delta-nested} documents exploratory
market-size and institutional-membership comparisons. The latter require
six-observation PID windows and vary with the comparison countries
(Figure~\ref{fig:accession-diagnostic}). Together these exercises provide
little support for an economic payoff or a common membership effect.
They do not identify governance differences or pre-existing economic
divisions as the explanation for the border pattern.

\FloatBarrier

\section{Discussion}
\label{sec:discussion}

The central empirical distinction is between the strength of a growth
relationship and the way information is shared across places. Domestic
city pairs tend to have stronger dependence, while domestic triplets
usually carry overlapping information. Cross-border pairs are less
dependent on average, yet their triplets more often have a synergistic
balance. These findings can coexist. As the Khagaria example shows, two
individually weak relationships can become informative when considered
jointly. Low bilateral dependence need not exhaust what a foreign city's
growth tells us in a wider geographic context.

\subsection{Context matters for interpreting a border relationship}

The PID comparison gives this distinction an economic-geographic reading.
A foreign city may repeat information already present in the target's
national environment, supply its own information, or add information through
its combination with that environment. Domestic comparison cities most
often have the largest unique component when they are on the target's
side. This locates the foreign connection relative to a specified context;
it does not label it intrinsically strong or weak. The choice of comparison
city is therefore part of the economic question, not simply a technical
control.

Related work uses embedding-based PID to evaluate a source's contribution
relative to a compressed representation of the surrounding system
\citep{hourican2026epid}. Here, the comparison context is an explicitly
selected city, allowing us to examine how the decomposition changes with
its geographic role.

Relational accounts emphasise how economic activity is embedded in
organisational and institutional connections \citep{bathelt2003,gluckler2007}.
Our findings offer a way to describe growth configurations consistent with
that emphasis, while remaining distinct from measures of the connections
themselves. Shared exposure can produce redundancy; combinations of
predictors can produce synergy. Neither information component identifies
trade, knowledge flows or complementary production. In particular, a
negative O-information score is not a quantity of output created by
cross-border cooperation. The weak growth associations and unstable
exploratory membership comparisons reinforce the need to keep information
structure separate from an economic mechanism.

\subsection{The geographic reference changes the question}

Map-wide mean O-information at ADM1, NUTS-1 and NUTS-2 lies within the
reference ranges from matched random subdivisions. This concerns the
mean within a specified random-map family, not equivalence of every local
relationship. It does not explain away the city-border contrast, because
international boundaries remain fixed. It does show why that contrast
cannot establish a general claim about administrative subdivisions.
The supplementary comparison between whole countries and an unmatched
random map illustrates how coverage and national grouping change the
object being measured.

This extends the measurement question raised by the MAUP literature
\citep{briant2010,dapena2016}: a geographic comparison must state which
features of the map are held constant. Matching typical region spacing
alone leaves coverage and national composition free to change. Conversely,
preserving countries deliberately removes international boundary placement
from the question. The reference map determines which claim the result
can support.

\subsection{Data and estimation limits}
\label{sec:discussion:caveats}

The most consequential limitation is the GDP product. National economic
accounts and nighttime-light allocation can generate shared city dynamics.
Growth differencing does not necessarily remove a time-varying national
component. Excluding countries with very limited internal variation preserves
the direction of the border contrast, but does not separate economic
connections from data construction in the remaining countries. Nighttime
lights are part of the input, so agreement with them would not be independent
validation. Replication with independently observed local accounts would
help establish how much of this geography reflects local economic activity.

The information measures also impose a specific statistical description.
Gaussian-copula estimates organise the correlations among transformed growth
series; they do not recover every form of nonlinear or tail dependence.
The scalar PID rule assigns zero unique information to the weaker source
within a triplet, while allowing it to contribute through synergy.
The large target-domestic unique component partly restates stronger
domestic pairwise dependence under this rule; the joint component describes
what the combination adds. The systematic sample uses one target orientation
per pair, so its frequencies do not establish a result for both orientations.

Finite-sample bias can make independent discrete variables appear
synergy-dominated under O-information \citep{gehlen2024bias}. That result
concerns discrete estimators; our Gaussian-copula estimates require their
own calibration. Twenty-seven annual observations give limited scope for
estimating uncertainty, especially with serial dependence. Calibration does
not support
individual-triplet discovery claims, and the six-observation institutional
windows are more restrictive still. Sensitivity checks address particular
choices rather than eliminating these limitations. Removing European crisis
years retains the direction of the border contrast but changes its magnitude;
using alternative domestic references changes border rankings. The study
therefore provides a description of geographic information patterns, with
explicit reference comparisons, rather than a ranking of causal border
barriers or an indicator of the benefits of integration.

\FloatBarrier

\section{Conclusion}
\label{sec:conclusion}

Studying three cities together distinguishes growth information that
repeats across places from information supplied by their combination.
In the gridded GDP estimates, negative O-information occurs in $29.93\%$
of cross-border triplets and $2.00\%$ of domestic triplets. This fifteenfold
contrast falls to approximately fourfold at comparable overall dependence,
and its direction persists under data-quality restrictions. Target-specific
decomposition shows how a foreign source can overlap with domestic
information or add through its combination with a comparison city.

The matched spatial comparison finds mean O-information under administrative
subdivisions within the range of random subdivisions inside the same
countries. It provides no basis for extending the city-border pattern into
a general result about administrative boundaries. The contribution is to
make two choices explicit when interpreting geographic growth dependence:
which cities are considered jointly, and which map defines the observations.
These choices organise information contained in growth correlations.
Establishing the underlying economic connections requires evidence beyond
the nationally allocated GDP estimates.

\FloatBarrier

\section{Data availability}
The underlying GDP product \citep{chen2022gdp} and GHSL functional urban
areas \citep{ghsfua2019} are available from their original providers.
The analysis code and derived results are available from the corresponding
author upon reasonable request. A public reproducibility archive will be
released upon journal publication.
Supplement~\ref{app:repro} describes the supporting code, data and validation records.

\ifdefined\AnonymousReview
\else
\section{Funding}
This work was supported by the Dutch Research Council (NWO) through the
SSH Open Competition XS programme [grant number 406.XS.25.01.061 to D.R.].

\fi

\section{Competing interests}
The authors declare no competing interests.

\section{Acknowledgements}
We used Claude (Anthropic) and Codex (OpenAI) to assist with coding, analysis
checks, figure preparation and manuscript revision. The authors directed
and reviewed this work and take responsibility for the final content.

\FloatBarrier
\bibliographystyle{oup-abbrvnat}
\bibliography{biblio}

\setcounter{section}{0}
\renewcommand{\thesection}{\Alph{section}}
\renewcommand{\theHsection}{supp.\Alph{section}}
\setcounter{figure}{0}
\renewcommand{\thefigure}{S\arabic{figure}}
\renewcommand{\theHfigure}{supp.\arabic{figure}}
\setcounter{table}{0}
\renewcommand{\thetable}{S\arabic{table}}
\renewcommand{\theHtable}{supp.\arabic{table}}
\section{Supplementary material}
\label{app:si}
\subsection{Data, geographic samples and exclusions}
\label{app:sample-construction}
\label{app:dispersion}

The city boundary dataset is
\texttt{GHS\_\allowbreak FUA\_\allowbreak UCDB2015\_\allowbreak GLOBE\_\allowbreak R2019A}
\citep{ghsfua2019}.

City GDP per capita divides annual polygon sums of the GDP raster by a
fixed population sum for each functional urban area (FUA). Population uses
GHS-POP R2023A, epoch 2015, at 1~km resolution in Mollweide coordinates
\citep{ghspop2023}. Because this denominator is fixed across 1992--2019,
the 27 temporal log changes equal changes in attributed total GDP.
GDP-weighted summaries use stored total-GDP sums. They do not incorporate
annual population change.

Table~\ref{tab:sample-rules} distinguishes the samples. Great-circle distances
use an Earth radius of 6,371~km. The city-triplet landscape uses polygon
centroids; distance bins and border maps use representative points inside
the polygons. The displayed distance bins contain 1,543,856 domestic and
404,590 cross-country pairs. These point conventions and geographic rules
are fixed within each comparison, rather than selected by its results.

\begin{table}[tbp]
\centering\small
\caption{Samples used in the figures. All information inputs are log
GDP-per-capita growth for 1993--2019 except the exploratory accession
windows, which contain six growth observations each. ``Restricted'' uses
the country/numerical rules with each analysis's own geographic eligibility.}
\label{tab:sample-rules}
\begin{tabular}{@{}p{.23\textwidth}p{.34\textwidth}p{.35\textwidth}@{}}
\toprule
Comparison & Geographic rule & Sample represented\\
\midrule
City triplets, Figure~\ref{fig:border-contrast} & All three centroid separations $\leq1,000$ km & Full 428,568,795; restricted 408,163,633 triplets\\
PID, Figure~\ref{fig:beta-atoms} & Cross-country source--target distance $\leq500$ km; contexts as below & Full 498,839; restricted 406,759 combinations\\
Distance, Figure~\ref{fig:distance-decay} & 50 km bands to 1,000 km & 1,948,446 restricted displayed pairs\\
Border maps, Figures~\ref{fig:border-map}, \ref{fig:border-map-raw}, \ref{fig:border-map-regions} & Pair distance $\leq300$ km, land adjacency and proximity rule & 87 borders, 19,809 restricted pairs\\
Matched regions, Figure~\ref{fig:partition-comparison} & All three centroid separations $\leq1,000$ km & Three scales; counts in Table~\ref{tab:partition-coverage}\\
City grouping, Figures~\ref{fig:spatial-partitions}, \ref{fig:spatial-partitions-minimum} & All three representative-point separations $\leq300$ km & Same 641 cities, six groupings\\
Country illustration, Figure~\ref{fig:partition-illustration} & All three centroid separations $\leq1,000$ km & 179 countries / 502 random regions; unmatched coverage\\
Accession, Figure~\ref{fig:accession-diagnostic} & Fixed eligible cross-country pairs and contexts & 75,731 pairs, four time windows\\
\bottomrule
\end{tabular}
\end{table}

\subsubsection{Selecting comparison cities for PID}

Each cross-country pair is enumerated once: the lower input row index is
the source and the higher is the target. This fixes a computational
orientation without using the outcome. The five candidate contexts are
the source's nearest domestic neighbour, the target's nearest domestic
neighbour, a capital-region representative from each country, and the
nearest third-country city within 1,000~km of the pair's coordinate midpoint.
Domestic neighbours minimise Euclidean distance in latitude and longitude;
pair and third-country cutoffs use great-circle distances. Capital matching
uses fixed names, aliases and regional representatives. If no match exists,
the country's most populous FUA in the dataset is used; the archive records
these substitutions.

Undefined contexts and contexts identical to an endpoint are omitted.
Different context categories can select the same city and remain distinct
rows. This gives 101,635 pairs and 498,839 combinations; restrictions retain
83,967 pairs and 406,759 combinations. Exclusions filter these fixed choices
without selecting replacement contexts. Each combination contributes once
to the pooled component frequencies, regardless of its information amount.

\subsubsection{Country variation and numerical eligibility}

The restrictions are a retrospective sensitivity analysis, declared before
the population comparisons after inspection had identified possible
near-copies. All 8,808 city series have 27 distinct annual observations,
yet 530 cities belong to 88 groups with identical rank paths. Temporal
variation therefore need not provide distinct information across cities.

For each country with at least two cities, calculate median growth across
cities at each date. For each city, calculate the median absolute deviation
from this national series across dates. The country statistic is the median
of those city deviations, with log growth multiplied by 100 to express
percentage points. The primary rule excludes countries at or below
$0.0001$ percentage points. Twenty single-city countries are unassessable
by this rule and are not classified as having zero dispersion.

The cutoff excludes 872 cities in 18 country/territory codes: BDI, CAF,
COD, COG, ETH, GNB, HKG, HTI, LBR, MAC, MDG, NER, NGA, PRK, RWA, SLE,
TCD and TGO. The 7,936 retained cities represent $99.05\%$ of sampled-city
2019 GDP. A low country median does not mean every city is identical.
The restriction changes geographic coverage; it neither establishes data
errors nor removes dependence on national accounting.

Separately, an eligible relationship must have minimum eigenvalue greater
than $10^{-6}$ in its unregularised normal-score correlation matrix.
For pairs this means $|\rho|<0.999999$. This rule addresses near-singularity,
not the sign or magnitude of the information score, and applies equally to
domestic and cross-country relationships. The primary screen combines both
rules. Table~\ref{tab:dispersion-alpha} separates them and varies their
thresholds; both worked examples pass every reported threshold.

\FloatBarrier

\subsection{Information measures, worked examples and inference limits}
\label{app:estimation}

For three variables, let $H$ denote entropy and $X_{-i}$ the pair excluding
$X_i$. Total correlation, dual total correlation and O-information are
\begin{align}
\mathrm{TC}&=\sum_{i=1}^3H(X_i)-H(X_1,X_2,X_3),\\
\mathrm{DTC}&=\sum_{i=1}^3H(X_{-i})-2H(X_1,X_2,X_3),\\
\label{eq:oinfo}
\Oinfo&=\mathrm{TC}-\mathrm{DTC}=\sum_{i<j}I(X_i;X_j)-\mathrm{TC}.
\end{align}
For triplets, O-information is co-information with this sign convention
\citep{rosas2019}. It describes a balance, not the total dependence.

For a series of length $n$, observations become normal scores
$z_t=\Phi^{-1}(r_t/(n+1))$, where $r_t$ is the ordinal rank and $\Phi$ the
standard-normal distribution function \citep{ince2017}. Ranks are recomputed
within each window; the city observations have no ties. Covariance uses
demeaned scores and divisor $n-1$, with diagonal ridge $10^{-10}$.
Figure~\ref{fig:city-examples}'s O-information annotations use the scalar
routine without a ridge. Rank normalisation changes marginal distributions
but does not remove non-Gaussian dependence or make levels equivalent to
growth.

For $d$ variables with covariance matrix $\Sigma$, Gaussian plug-in entropy is
$H_d=\tfrac12\log[(2\pi e)^d\det\Sigma]$. The landscape and pairwise
measures subtract the finite-sample term
\begin{equation}
b(d,n)=\frac12\left[\sum_{i=1}^{d}\psi\left(\frac{n-i}{2}\right)
                 -d\log\left(\frac{n-1}{2}\right)\right],
\end{equation}
where $\psi$ is the digamma function. With correlation matrix $R$ and
pairwise correlations $\rho_{ij}$, the plug-in formulas are
\begin{equation}
I(X_i;X_j)=-\tfrac12\log(1-\rho_{ij}^2),\qquad
\Oinfo=\tfrac12\left[\log\det R-\sum_{i<j}\log(1-\rho_{ij}^2)\right].
\end{equation}
The correction adds $-3b(1,n)+3b(2,n)-b(3,n)$ to O-information.
Estimated MI can become slightly negative, although population MI is
non-negative. These Gaussian corrections do not guarantee unbiased
estimation for rank-transformed, serially dependent growth.

\subsubsection{Partial information decomposition}

For target $Y$ and sources $S_1,S_2$, the PID identities are
\begin{align}
\label{eq:pid}
I(Y;S_1,S_2)&=R+U_1+U_2+S,\\
I(Y;S_1)&=R+U_1,\qquad I(Y;S_2)=R+U_2,
\end{align}
where $R$ is redundancy, $U_j$ unique information and $S$ synergy
\citep{williams2010,bertschinger2014}. We use plug-in Gaussian information
without the entropy correction. For these scalar variables, the deficiency
calculation agrees with the minimum-mutual-information rule
\citep{barrett2015,venkatesh2022}:
\begin{align}
R&=\min\{I(Y;S_1),I(Y;S_2)\},\\
U_j&=I(Y;S_j)-R,\\
S&=I(Y;S_1,S_2)-\max\{I(Y;S_1),I(Y;S_2)\}.
\end{align}
The weaker source has zero unique information within each triplet; either
source can be stronger across triplets. Co-information equals $R-S$ when
calculated with the same conventions. It therefore differs slightly from
the corrected O-information displayed here. Independent high-precision
checks confirm the implementation on all 498,839 decompositions; the
validation archive contains the numerical comparisons. They do not
validate the Gaussian copula or all possible PID definitions.

\subsubsection{The two city examples}
\label{app:examples}

The Thai example uses Nakhon Pathom (eFUA 1724), Bangkok (1894) and target
Lopburi (2048); the cross-border example uses Hetauda (1327) and Itahari
(2033), Nepal, with target Khagaria (10198), India. Separate and joint
raw-growth fits have $R^2$ values $0.9055$, $0.9357$, $0.9371$ and
$0.1046$, $0.0228$, $0.6597$, respectively. All are ordinary least squares
with an intercept. The joint models have two additive terms, with no
product, lag or interaction:
\begin{align*}
\widehat g_{\rm Lopburi}&=-0.002411+0.186869g_{\rm Nakhon\ Pathom}
                       +0.962005g_{\rm Bangkok},\\
\widehat g_{\rm Khagaria}&=0.063180+1.563518g_{\rm Hetauda}
                        -1.887214g_{\rm Itahari}.
\end{align*}
Growth is in decimal log units. The Nepali sources have raw correlation
$0.8274$; their opposite-signed coefficients exploit a contrast between
them. The complete covariance matrix determines these linear fits.

\begin{table}[tbp]
\centering
\caption{PID of the two worked examples, in nats. Source ordering follows
Figure~\ref{fig:beta-atoms}; values are rounded independently.}
\label{tab:example-pid}
\begin{tabular}{@{}lrr@{}}
\toprule
Component & Lopburi & Khagaria\\
\midrule
Redundancy & 1.972060 & 0.000822\\
Source 1 unique & 0.000000 & 0.000000\\
Source 2 unique & 0.274281 & 0.068722\\
Synergy & 0.001975 & 0.204942\\
Joint information & 2.248316 & 0.274486\\
\bottomrule
\end{tabular}
\end{table}

Table~\ref{tab:example-pid} instead uses normal-score observations. Its
$R-S$ values are approximately $1.970084$ and $-0.204120$ nats, compared
with corrected O-information $1.970952$ and $-0.203253$. For Khagaria,
Hetauda and Itahari supply $0.000822$ and $0.069544$ nats separately,
although Hetauda has the larger raw-growth $R^2$. PID does not partition
that raw $R^2$. The examples were deliberately selected for contrasting
configurations, unambiguous identities, sub-national variation and a
regression gain present in raw growth. They illustrate the measures rather
than test a hypothesis.

\subsubsection{Why the results are descriptive}
\label{app:numit}

NuMIT varies information's arrangement at a fixed total: joint target
information for PID and TC for O-information \citep{liardi2025}. With
1,000 reference draws, the implemented two-sided empirical probability has
minimum $2/1001$, approximately 0.002. The first Benjamini--Hochberg
threshold at $q=0.05$ for 498,839 tests is about $10^{-7}$
\citep{benjamini1995}. Later thresholds can admit sufficiently many small
probabilities, but isolated extremes cannot be resolved. Overlapping
triplets also complicate the procedure's dependence assumptions.

Table~\ref{tab:calibration} compares conditioning on estimated versus known
generating information in independent Gaussian systems with 27 observations.
PID uses 200 systems per setting and 1,000 reference draws; O-information
uses 1,000 systems and a shared reference table. Estimating the conditioning
quantity fails to reproduce the nominal 5\% rejection rate across settings.

\begin{table}[tbp]
\centering
\caption{Gaussian calibration at 27 observations. Entries are rejection
percentages under the stated reference family. Estimated-information
conditioning does not reproduce nominal calibration across these settings.}
\label{tab:calibration}
\begin{tabular}{@{}llrr@{}}
\toprule
Measure & True information (nats) & Estimated conditioning & Known conditioning\\
\midrule
PID synergy & 0.01 & 3.0 & 4.5\\
 & 0.1 & 1.5 & 3.5\\
 & 1 & 9.5 & 5.5\\
 & 3 & 15.5 & 6.0\\
O-information & 0.01 & 0.0 & 5.5\\
 & 0.1 & 0.1 & 5.8\\
 & 0.5 & 1.3 & 5.5\\
 & 1.5 & 12.7 & 5.8\\
\bottomrule
\end{tabular}
\end{table}

The table excludes a higher-TC diagnostic that failed its numerical
agreement criterion; the complete diagnostics and sample-length simulations
are in the validation archive. These independent-observation experiments
do not calibrate the economic time series: city normal-score lag-one
correlations have median 0.378 and interquartile range 0.188--0.555.
They do not identify an effective sample size. We make no individual-triplet
or component significance claims, and non-rejection is not evidence of
absent synergy. Tail extrapolation at the required extreme probabilities
has not been validated.

\FloatBarrier

\subsection{Additional city results and robustness}
\label{app:corrected-results}

Table~\ref{tab:corrected-alpha} compares all eligible city triplets with
the primary restricted sample. Each group's mean and negative-score
fraction use its own denominator; the overall fraction weights groups by
their actual triplet counts. A negative score describes a synergistic
balance under the corrected estimator, not a significant detection.

\begin{table}[tbp]
\centering
\caption{O-information of city GDP-per-capita log growth. Negative estimated scores
indicate a synergistic balance under the finite-sample convention in
Supplement~\ref{app:estimation}. Both samples retain their own
group denominators; the primary screen applies the country-dispersion
and numerical-eligibility rules.}
\label{tab:corrected-alpha}
\begin{tabular}{@{}lrrr@{}}
\toprule
Sample & Triplets & Mean $\Oinfo$ (nats) & Negative (\%)\\
\midrule
\multicolumn{4}{@{}l}{Full reference}\\
All & 428,568,795 & 0.205920 & 7.1466\\
Within-country & 349,632,590 & 0.244741 & 2.0031\\
Cross-border & 78,936,205 & 0.033972 & 29.9287\\
\addlinespace
\multicolumn{4}{@{}l}{Primary screen}\\
All & 408,163,633 & 0.201853 & 6.1836\\
Within-country & 340,852,271 & 0.234150 & 2.0115\\
Cross-border & 67,311,362 & 0.038303 & 27.3106\\
\bottomrule
\end{tabular}
\end{table}

The full-reference cross-border/within-country ratio is $14.9412$;
restriction gives $13.5774$. To compare similar dependence strengths,
we use 100 pooled full-reference TC quantile bins, requiring at least
1,000 triplets per group per bin. The edges remain fixed across samples.
Within each sample, cross-border counts supply weights applied to both
groups; all bins have support, including bins with no negative scores.
The standardised rates are $29.9287\%$ versus $7.4772\%$ in the full
reference (ratio $4.0027$), and $27.3106\%$ versus $7.5522\%$ after
restriction (ratio $3.6163$). TC matching does not match income, geography,
industrial structure or national growth factors.

\begin{table}[tbp]
\centering
\small
\caption{Global negative-O fractions under declared restrictions. Each
row filters the original geographic triplets. Full and primary results are in Table~\ref{tab:corrected-alpha}.
Single-rule rows apply only the named rule; other rows vary one cutoff
with the other fixed at its primary value. The TC ratio uses fixed full-reference quantile edges for all rows.}
\label{tab:dispersion-alpha}
\begin{tabular}{@{}lrrrrr@{}}
\toprule
Restriction & Triplets & Within (\%) & Cross (\%) & Raw ratio & TC ratio\\
\midrule
Country only & 408,320,803 & 2.0111 & 27.2731 & 13.561 & 3.616\\
Numerical only & 427,410,010 & 2.0078 & 30.0592 & 14.971 & 4.003\\
Country cutoff $10^{-5}$ pp & 423,187,061 & 2.0086 & 29.9735 & 14.923 & 3.969\\
Country cutoff $10^{-3}$ pp & 407,893,012 & 2.0115 & 27.1878 & 13.516 & 3.607\\
Eigenvalue cutoff $10^{-8}$ & 408,163,633 & 2.0115 & 27.3106 & 13.577 & 3.616\\
Eigenvalue cutoff $10^{-4}$ & 408,163,632 & 2.0115 & 27.3106 & 13.577 & 3.616\\
\bottomrule
\end{tabular}
\end{table}

Table~\ref{tab:dispersion-alpha} separates the two restrictions and varies
one threshold at a time. The cross-border contrast survives these choices.
Thresholds address data variation and covariance conditioning, without
selecting on negative information scores.

\subsubsection{PID amounts and component frequencies}

Table~\ref{tab:corrected-beta} retains the same context choices when
filtering rows. Every component mean uses the same rows within its sample;
the unrounded means sum to joint information, $0.499296$ nats in the full
reference and $0.501252$ in the restricted sample. A dominant-component
frequency counts combinations in which that component is largest; it is
not its share of information. Adding the foreign source after observing
the context provides source-unique information plus synergy,
$I(Y;S_1\mid S_2)=U_1+S$.

\begin{table}[tbp]
\centering\small
\caption{PID amounts and dominant-component frequencies. Full and restricted
samples contain 498,839 and 406,759 combinations, respectively. Means are
in nats; frequencies give percentages of combinations. Components are
rounded independently.}
\label{tab:corrected-beta}
\label{tab:dispersion-pid}
\begin{tabular}{@{}lrrrr@{}}
\toprule
 & \multicolumn{2}{c}{Full reference} & \multicolumn{2}{c}{Primary screen}\\
Component & Mean & Dominant (\%) & Mean & Dominant (\%)\\
\midrule
Context-unique & 0.384484 & 50.7506 & 0.378319 & 50.1707\\
Redundancy & 0.076611 & 26.2947 & 0.084196 & 27.6857\\
Source-unique & 0.014242 & 12.6911 & 0.014905 & 12.8162\\
Synergy & 0.023959 & 10.2636 & 0.023831 & 9.3274\\
\bottomrule
\end{tabular}
\end{table}

\subsubsection{GDP-weighted participation}
\label{app:gdp-participation}

Let $N_i$ count eligible triplets containing city $i$, and $N_i^-$ those
with $\Oinfo<0$. For $N_i>0$, let $q_i=N_i^-/N_i$. With GDP $G_{it}$,
\begin{equation}
Q_t=\frac{\sum_{i:N_i>0}G_{it}q_i}{\sum_{i:N_i>0}G_{it}}.
\end{equation}
This selects a city in proportion to GDP, then an eligible triplet
containing it. Each city's GDP enters once. It measures participation,
not output generated by synergy.

In the full reference, 8,797 cities have eligible triplets and cover
$99.8841\%$ of sampled-city 2019 GDP; $Q_{2019}=7.2119\%$. Of these,
8,294 participate in at least one negative-score triplet, accounting for
$88.6155\%$ of eligible-city GDP. After restriction, 7,924 cities remain
eligible, covering $98.9297\%$ of original sampled-city GDP.
$Q_{2019}=7.1766\%$; 7,417 have at least one negative-score triplet and
account for $88.5077\%$ of restricted eligible-city GDP. The denominators
are sampled-city GDP, not world output. The archive reports alternative
years and equal-city weights.

\subsubsection{Sensitivity to crisis years}
\label{app:robustness}

The crisis comparison uses 601 European cities, with 347,117 within-country
and 5,553,625 cross-border triplets whose separations are all within
1,000~km. The panel agrees with the corresponding global observations.
We remove 2008--09, 2011--12 or both pairs of growth observations, recomputing
ranks and sample-length corrections while holding geography fixed.
These are European comparisons, not a global crisis test.

\begin{table}[tbp]
\centering\small
\caption{European crisis exclusions: negative-O percentages before and after
TC standardisation. $T$ counts growth observations. Ratios divide cross-border
by within-country rates. Coverage is the percentage of each original group
retained in common TC support; raw rows use the complete geographic sample.}
\label{tab:crisis-raw}
\label{tab:crisis-matched}
\begin{tabular}{@{}lrrrrrr@{}}
\toprule
 & & & & & \multicolumn{2}{c}{Coverage (\%)}\\
Scenario & $T$ & Cross (\%) & Within (\%) & Ratio & Cross & Within\\
\midrule
\multicolumn{7}{@{}l}{Unstandardised}\\
Full period & 27 & 9.0200 & 0.0173 & 521.8 & 100 & 100\\
Drop 2008--09 & 25 & 22.6052 & 0.1253 & 180.4 & 100 & 100\\
Drop 2011--12 & 25 & 8.1064 & 0.0723 & 112.1 & 100 & 100\\
Drop both & 23 & 22.1503 & 0.3702 & 59.8 & 100 & 100\\
\addlinespace
\multicolumn{7}{@{}l}{Standardised to cross-border TC weights on common support}\\
Full period & 27 & 7.5907 & 0.00457 & 1661.8 & 39.45 & 99.81\\
Drop 2008--09 & 25 & 17.9435 & 0.05629 & 318.8 & 40.51 & 99.77\\
Drop 2011--12 & 25 & 7.5579 & 0.02621 & 288.3 & 39.47 & 99.47\\
Drop both & 23 & 18.9716 & 0.25437 & 74.6 & 41.59 & 99.54\\
\bottomrule
\end{tabular}
\end{table}

For each scenario, we estimate 100 pooled TC quantile bins, retaining bins
with at least 500 triplets per group. Cross-border bin counts weight both
groups; zero-negative-score bins remain eligible. The four scenarios retain
43, 44, 43 and 45 bins. They do not represent one fixed TC distribution.
The contrast survives, but its ratio is sensitive to the small domestic
negative fraction: the raw domestic numerator rises from 60 to 1,285
triplets when both crisis pairs are removed. Meanwhile the matched ratio
falls from 1,661.8 to 74.6 and the rate difference rises from 7.59 to
18.72 percentage points. Year deletion changes ranks, sample length and
weights; these changes cannot isolate the contribution of common shocks.

Requiring numerical eligibility in every period retains a common cohort
of 346,419 domestic and 5,550,486 cross-border triplets. All European
countries pass the country rule. Table~\ref{tab:crisis-screened} re-estimates
bins on this cohort under the same support rule. Although few triplets
are removed, losing a high-TC bin reduces domestic matching coverage to
about 83\%. The direction persists, but the represented population differs
from the full-reference standardisation.

\begin{table}[tbp]
\centering
\small
\caption{Crisis exclusions on the common numerically eligible cohort.
The last two columns give common-support coverage of each screened group.}
\label{tab:crisis-screened}
\begin{tabular}{@{}lrrrrrr@{}}
\toprule
Scenario & Raw ratio & \shortstack{Matched\\cross (\%)} & \shortstack{Matched\\within (\%)} & \shortstack{Matched\\ratio} & \shortstack{Cross\\coverage (\%)} & \shortstack{Within\\coverage (\%)}\\
\midrule
Full period & 521.0 & 7.5962 & 0.00457 & 1661.9 & 39.45 & 82.80\\
Drop 2008--09 & 180.1 & 17.9535 & 0.05593 & 321.0 & 40.52 & 82.79\\
Drop 2011--12 & 111.9 & 7.5671 & 0.02638 & 286.9 & 39.47 & 82.49\\
Drop both & 59.7 & 18.9936 & 0.25480 & 74.5 & 41.59 & 82.59\\
\bottomrule
\end{tabular}
\end{table}

\subsubsection{Estimator sensitivity}

Nearly copied city series can produce ridge-sensitive information amounts.
In a deliberately selected full-reference diagnostic, the 32 highest-information
PID cases change by about 2.30 nats in at least one component when the ridge
changes by a factor of 100. This is not a prevalence estimate. Across the
complete restricted PID sample, ridges zero, $10^{-12}$, $10^{-10}$ and
$10^{-8}$ change no dominant component and no amount by more than
$1.61\times10^{-5}$ nats. Thus numerical correctness alone does not remove
the data limitation. Neither this check nor rank normalisation validates
nonlinear or tail dependence beyond the Gaussian copula, or inference under
temporal dependence. Detailed solver and selected-case checks are archived.

\FloatBarrier

\subsection{Border comparisons and supporting maps}
\label{app:data-quality}
\label{app:border-sensitivity}
\label{app:supporting-figures}

\subsubsection{Raw gaps and the cohesion adjustment}

The primary maps use restricted pairs at most 300~km apart in countries
sharing a land border. The shared boundary must be no farther from the
pair's coordinate midpoint than the cities' separation. Boundary distance
uses a per-border azimuthal-equidistant projection. This proximity rule
does not require the line between the cities to cross the border.
Each border needs at least 30 pairs. For each pair, the raw gap is pooled
domestic mean MI in its 50~km band minus that pair's MI. Domestic bands
need at least 30 pairs; references are recomputed after restriction.

Let $C_a,C_b$ be each country's mean domestic MI in the original 1,000~km
neighbourhood. For border mean gap $D_{ab}$, fit
\begin{equation}
D_{ab}=\alpha+\beta_L\min(C_a,C_b)+\beta_H\max(C_a,C_b)+e_{ab}.
\end{equation}
Each border has equal weight. A positive residual means a larger raw gap
than predicted from domestic cohesion; a negative residual means a smaller
gap, which can still be positive. Raw and adjusted maps share 87 borders,
19,809 pairs and 2,704 cities. Germany--Luxembourg lacks Luxembourg cohesion
and is excluded from both. The fit has $R^2=0.332$; 84 raw border means
are positive. Neither residual signs nor this regression identify causal
border effects.

Restriction removes 21 borders from the full-reference adjusted sample.
Among the 87 retained borders, full/restricted residuals correlate at
$0.9900$; Belgium--Netherlands alone changes sign. The archive lists
exclusions by country rule, pair support and missing cohesion.

\subsubsection{Distance and domestic-reference sensitivity}

Table~\ref{tab:border-caps} varies the cap, holding 50~km bands, screening,
adjacency, proximity and minimum pair support fixed. Proximity is evaluated
through 400~km. Domestic references and cohesion use the same screened
domestic universe; each model is refitted on its supported borders.

\begin{table}[tbp]
\centering\small
\caption{Sensitivity of border gaps. Panel (a) varies the distance cap with
the pooled domestic reference. Panel (b) changes the reference at 300~km;
its last three rows use identical pairs and borders. Positive counts refer
to raw border means. Mean gap averages borders equally, in nats.}
\label{tab:border-caps}
\label{tab:border-references}
\begin{tabular}{@{}lrrrrr@{}}
\toprule
\multicolumn{6}{@{}l}{(a) Distance limit}\\
Cap (km) & Borders & Pairs & Positive & Negative & $R^2$\\
\midrule
200 & 38 & 6,762 & 36 & 2 & 0.450\\
300 & 87 & 19,809 & 84 & 3 & 0.332\\
400 & 121 & 40,056 & 117 & 4 & 0.274\\
\addlinespace
\multicolumn{6}{@{}l}{(b) Domestic reference at 300 km}\\
Reference & Borders & Pairs & Positive & Mean gap & $R^2$\\
\midrule
Pooled, native support & 87 & 19,809 & 84 & 0.356 & 0.332\\
Pooled, bilateral support & 30 & 12,949 & 28 & 0.355 & 0.483\\
Equal-country & 30 & 12,949 & 30 & 0.950 & 0.966\\
Two-country pair-weighted & 30 & 12,949 & 30 & 0.899 & 0.880\\
\bottomrule
\end{tabular}
\end{table}

On the 38 borders supported at all caps, models refitted to this common
set have adjusted-gap rank correlations of $0.9937$ (200~km) and $0.9963$
(400~km) with the 300~km result. No residual changes sign at 200~km; two do
at 400~km. The cap changes coverage more than ordering, although the pair
composition within these borders still changes.

For countries $a,b$ and band $d$, let $B_{a,d},B_{b,d}$ be domestic means
with counts $n_{a,d},n_{b,d}$. The equal-country reference is
$(B_{a,d}+B_{b,d})/2$; the pair-weighted reference is
$(n_{a,d}B_{a,d}+n_{b,d}B_{b,d})/(n_{a,d}+n_{b,d})$.
Each country must supply 30 domestic pairs in that band, without filling
missing bands. This leaves 30 borders and 12,949 pairs at 300~km. We also
recalculate the pooled reference on this identical support. Domestic pairs
need not lie near the border, and these comparisons cannot represent
countries lacking supported bands.

All 30 raw border means are positive under both bilateral averages and
each country's separate reference. Yet raw rankings correlate with pooled
rankings at only $-0.1818$ and $-0.2974$. After refitting cohesion, the
correlations are $0.9528$ and $0.6218$, with two and seven residual sign
changes. Equal-country adjustment is relatively stable, but neither map is
invariant to reference weighting. Its high $R^2$ is partly structural:
the baseline and regressors both use domestic MI.

The finite-sample MI correction at 27 observations is about $-0.0204$
nats. Slightly negative estimates are retained; the same additive constant
cancels in domestic-minus-cross-country gaps. An estimated gap can therefore
slightly exceed its estimated domestic baseline. It measures relative
dependence, not higher-order synergy. Computational bounds are documented
in the archive.

\begin{figure}[htbp]
\centering
\includegraphics[width=\textwidth,height=.58\textheight,keepaspectratio]{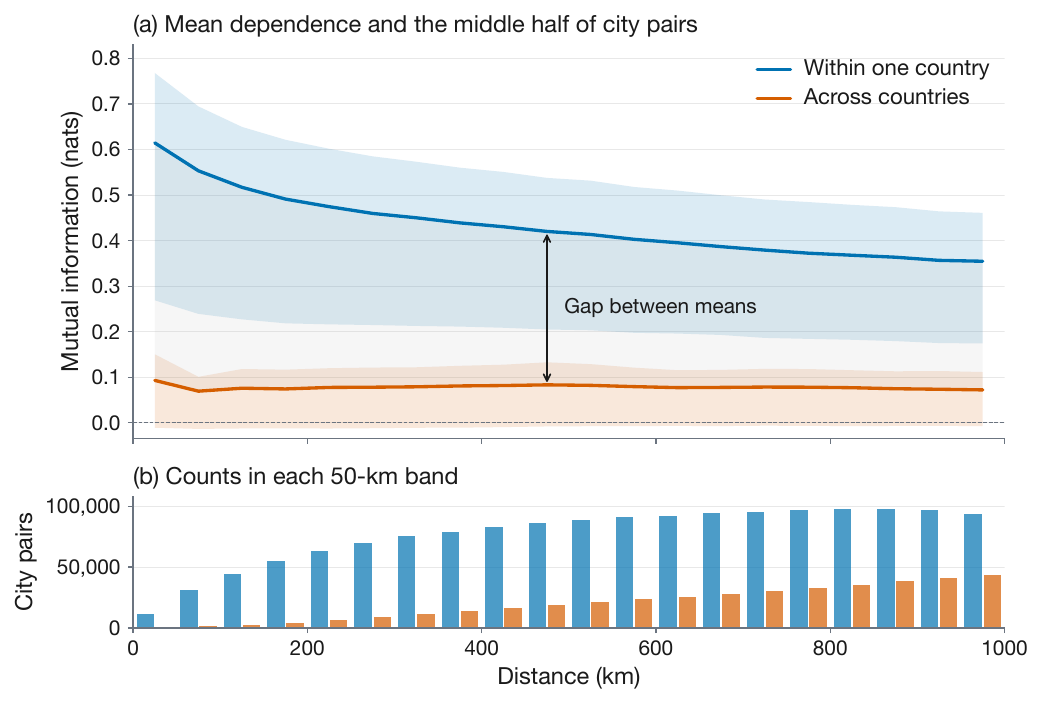}
\caption{Dependence between city GDP-per-capita growth series at different distances.
Lines show mean Gaussian-copula mutual information, in nats, in 50~km
distance bands for same-country and cross-country pairs, using annual log
growth for 1993--2019. Higher values indicate stronger dependence, including
either positive or negative association between the growth series.
After the country data-quality and numerical exclusions described in the
Methods, the displayed bins contain 1,948,446 pairs.
Blue and orange ribbons show the interquartile range across pairs; the
pale fill between the lines and the double arrow mark the gap between
group means. These ribbons describe variation across pairs.
The lower panel shows separate blue and orange count bars for the two
groups in each distance band. Within-country dependence is higher throughout
the displayed distance range. This pooled comparison holds distance bands,
but not country composition or other economic characteristics, comparable.}
\label{fig:distance-decay}
\end{figure}

\begin{figure}[htbp]
\centering
\includegraphics[width=\textwidth,height=.58\textheight,keepaspectratio]{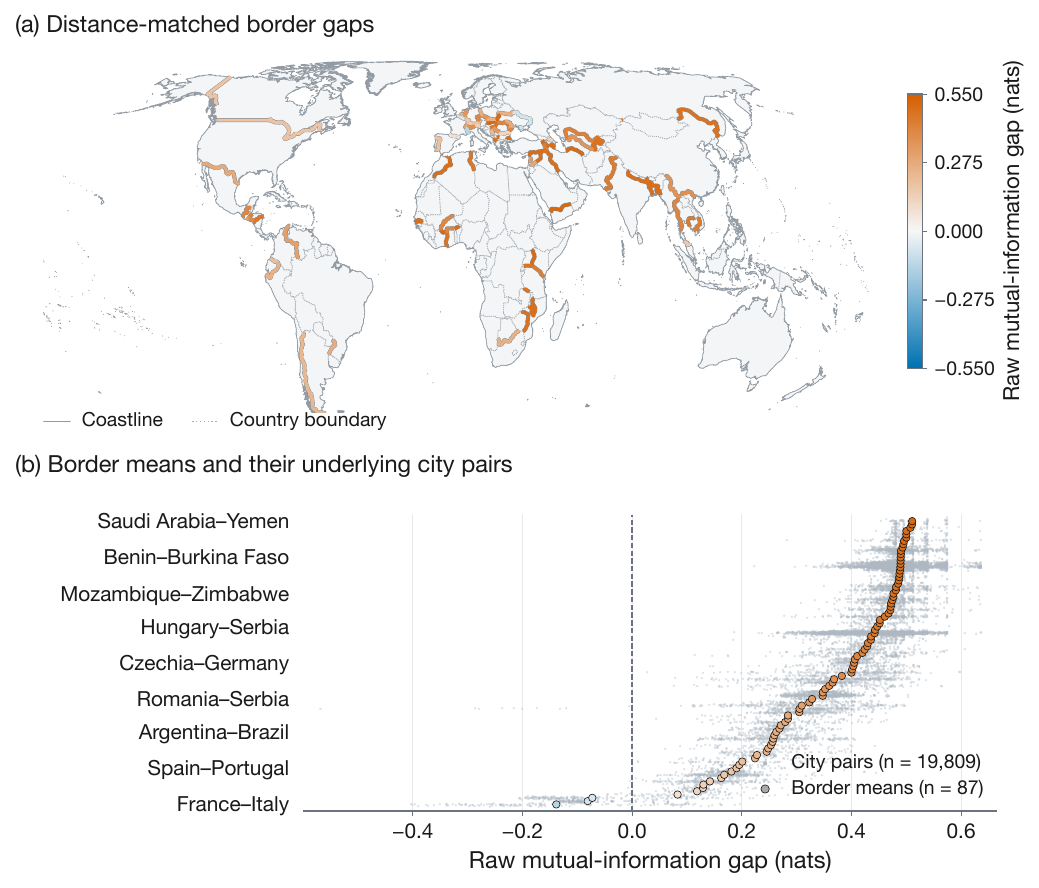}
\caption{Geographic variation in the gap in city growth dependence.
For each qualifying cross-country pair (cities at most 300~km apart in
countries sharing a land border, with the shared border no farther from
their midpoint than the cities are from one another), the gap is the pooled
within-country mean mutual information in its 50~km distance band minus
the pair's mutual information. Mutual information measures dependence
between annual GDP-per-capita log-growth series for 1993--2019.
The data-quality exclusions remove countries with very limited spatial
variation in city growth and pairs with numerically unstable estimates, as
described in the Methods.
Panel (a) colours 87 shared land borders by their mean gap.
Panel (b) ranks the same means and shows the 19,809 underlying pair gaps
in grey. Positive values indicate lower dependence than the pooled
domestic reference; negative values indicate higher dependence.
Of the 87 border means, 84 are positive. Each displayed border has at least
30 retained pairs and domestic-cohesion data, so the sample matches
Figure~\ref{fig:border-map}; domestic distance baselines are recomputed
after the exclusions. Grey coastlines and dotted boundaries provide
context. Thin outlines keep eligible borders visible near zero;
dotted-only borders are outside the sample. The colour scale spans
$-0.55$ to $+0.55$ nats without clipping any border mean.
Sensitivity to the domestic reference is reported in
Supplement~\ref{app:border-sensitivity}.}
\label{fig:border-map-raw}
\end{figure}

\begin{figure}[htbp]
\centering
\includegraphics[width=\textwidth,height=.58\textheight,keepaspectratio]{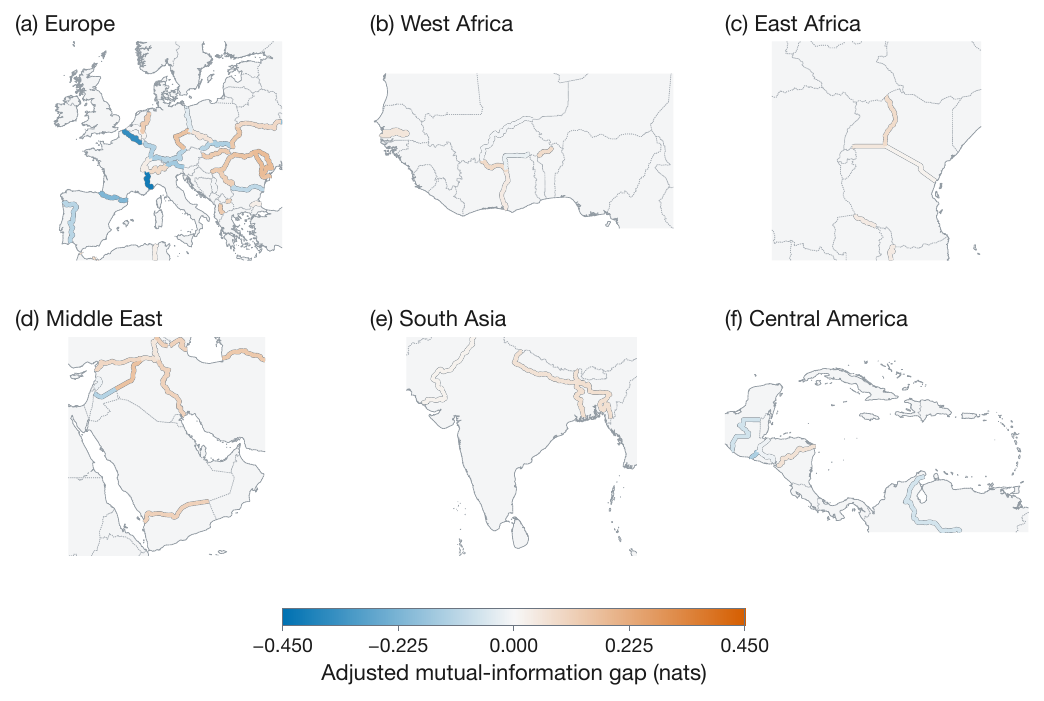}
\caption{Regional enlargements of the cohesion-adjusted border map.
  Panels show Europe, West Africa, East Africa, the Middle East, South Asia
  and Central America. All measures use Gaussian-copula mutual information between annual
  GDP-per-capita log-growth series for 1993--2019. Values are the same
  border-level residuals as in
  Figure~\ref{fig:border-map}, using its $-0.45$ to $+0.45$ nat
  colour scale without clipping in any panel.
  Positive values indicate a raw mutual-information gap larger than predicted
  by the adjoining countries' domestic cohesion; negative values indicate
  a smaller gap. No regional regression or rescaling is applied. Coastlines
  and dotted country boundaries show location. Thin outlines identify
  eligible borders, including those with residuals near zero; dotted-only
  boundaries are outside the eligible screened sample. The source is the same 87-border
  refit as the main map; these panels do not estimate separate regional fits.}
\label{fig:border-map-regions}
\end{figure}

\FloatBarrier

\subsection{Spatial grouping and matched partitions}
\label{app:city-grouping}
\label{app:maup-design}
\label{app:nuts3-description}

\subsubsection{The same cities under six groupings}

\begin{figure}[htbp]
\centering
\includegraphics[width=\textwidth,height=.58\textheight,keepaspectratio]{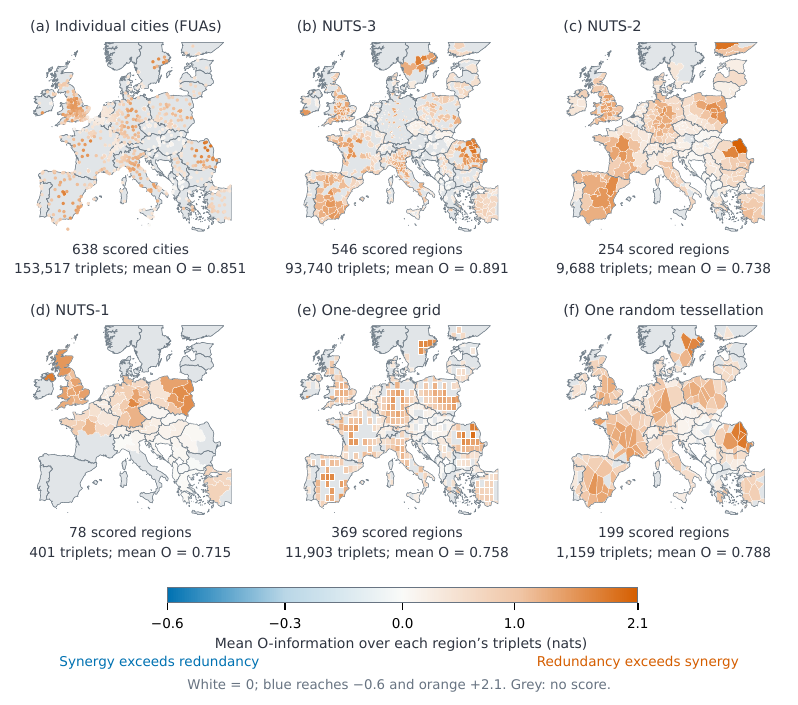}
\caption{\textbf{The same 641 cities, grouped six ways.}
All panels use the same cities in 34 countries within
$11^{\circ}$W--$32^{\circ}$E and $35^{\circ}$--$62^{\circ}$N,
restricted to the common European footprint. Panels (a--f) show individual
functional urban areas (FUAs), NUTS-3, NUTS-2, NUTS-1, a one-degree grid
and one random tessellation. Each city contributes its annual
GDP-per-capita log-growth series for 1993--2019. Within each occupied
region, these series are averaged with equal city weights. Gaussian-copula
O-information is then estimated for triplets of regions whose three
separations are all at most 300~km, applying the data-quality and numerical
exclusions described in the Methods.
Each city's or region's colour shows the mean O-information over the
eligible triplets containing it. Orange indicates a redundancy-dominated
mean; blue indicates a synergy-dominated mean. Grey means no eligible score.
The common colour scale runs from $-0.6$ to $+2.1$ nats, with white at zero.
Colour intensity is scaled separately below and above zero; no values are clipped.
City points in (a) aid visibility. Panel labels give the number of scored
cities or regions, the number of eligible triplets and the mean across
those triplets. The random map can cross country boundaries and illustrates
an alternative grouping. Unlike Figure~\ref{fig:partition-comparison},
these panels average city growth rather than aggregate raster GDP and
population. These maps describe sensitivity to spatial grouping.}
\label{fig:spatial-partitions}
\end{figure}

\begin{figure}[htbp]
\centering
\includegraphics[width=\textwidth,height=.58\textheight,keepaspectratio]{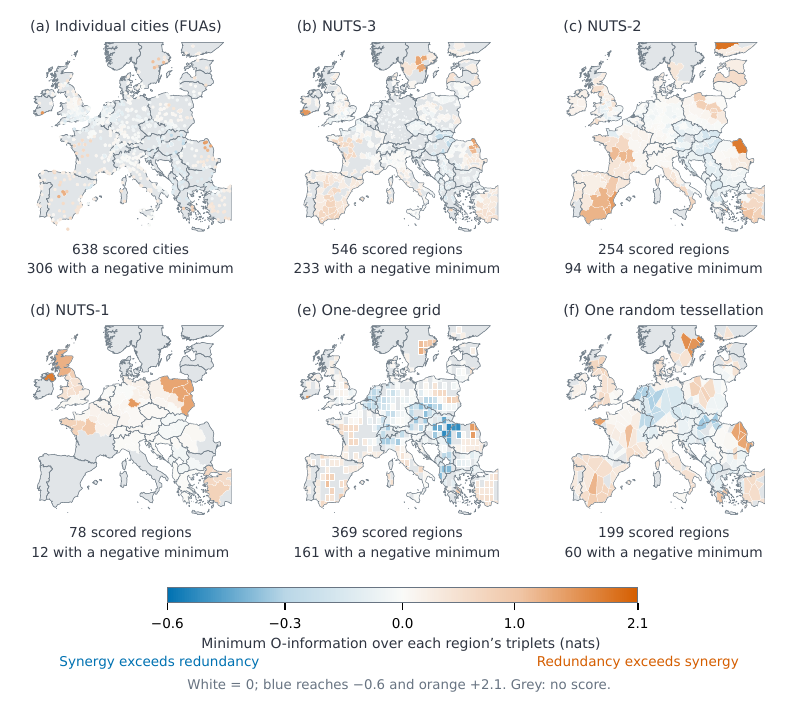}
\caption{\textbf{The most synergy-dominated triplet involving each city or region.}
This additional version uses exactly the same cities, groupings, growth
series and eligible triplets as Figure~\ref{fig:spatial-partitions}.
Each city's or region's colour now shows the \emph{minimum} O-information
among the eligible triplets containing it. A blue region therefore
participates in at least one triplet with negative estimated O-information, even if its mean across triplets is positive. Orange means
that even the minimum is positive. Grey means no eligible score.
Both versions share a $-0.6$ to $+2.1$ nat colour scale, with white at zero.
Colour intensity is scaled separately below and above zero; no values are clipped.
Panel labels count scored cities or regions and those with a negative
minimum. In (a), 306 of 638 scored cities participate in a negative-O
triplet, although none has a negative mean. The minimum selects an extreme;
it tends to decrease as a region participates in more triplets. It describes
the strongest synergy-dominated balance found, rather than its prevalence
or statistical significance.}
\label{fig:spatial-partitions-minimum}
\end{figure}

Figures~\ref{fig:spatial-partitions} and~\ref{fig:spatial-partitions-minimum}
fix membership before comparing groupings. Of 791 city points in the
European window, 641 match the common footprint and every grouping; the
other 150 are excluded throughout, without nearest-region imputation.
NUTS assignments require the same country. Each occupied region receives
the unweighted mean of its cities' 27 growth observations before rank
normalisation. This differs from aggregating GDP and population rasters
in the matched comparison below.

Triplets require all three representative-point separations within 300~km
and minimum normal-score correlation eigenvalue above $10^{-6}$.
The Gaussian correction and ridge follow Supplement~\ref{app:estimation}.
Fixing city membership does not fix unit counts or eligible triplets.
The illustrative random map uses uniform longitude/latitude seeds, without
population weighting, country containment or matching gates.

A region's mean averages eligible triplets containing it; its minimum
selects the most negative. Positive means can coexist with negative minima.
Across the six groupings, negative means occur in 0, 0, 1, 2, 1 and 3 units;
negative minima occur in 306, 233, 94, 12, 161 and 60. The lowest mean and
minimum are $-0.0404$ and $-0.5213$ nats. More triplets give more opportunities
for an extreme minimum, so this is not a prevalence or significance measure.
Panel-wide means weight triplets equally rather than averaging regional
colours. Assignment, occupancy and spacing tables are archived.

\FloatBarrier

\subsubsection{Observed regions and common geographic coverage}

The matched comparison uses geoBoundaries ADM1 and GISCO 2016 NUTS regions.
Provider geometries are clipped to fixed country ownership. Taiwan has a
separate mask; Hong Kong and Macao retain distinct labels, and Palestine
uses its provider regions. These attribution conventions avoid overlap.
ADM1 has 2,880 regions with positive population and GDP throughout and
finite nonconstant growth, covering 8,702 city reference points. Country
restrictions retain 2,654 regions in 163 codes. Uncovered points are not
filled from neighbours. NUTS-1 and NUTS-2 use the intersection of their
occupied native raster pixels with agreeing country ownership, giving
identical coverage, including source overseas pieces.

Regional series sum each year's GDP cells, divide by fixed 2015 population
and take 27 log changes. Each native pixel is assigned once. Observed and
random maps have identical country ownership and raster support.
Table~\ref{tab:partition-coverage} reports primary eligible samples.

\begin{table}[tbp]
\centering\small
\caption{Observed regional samples. Triplets require three centroid distances
at most 1,000~km and minimum correlation eigenvalue above $10^{-6}$.
ADM1, NUTS-1 and NUTS-2 have matched random references; the final row is
descriptive only. Mean O-information is in nats.}
\label{tab:partition-coverage}
\label{tab:nuts3-description}
\begin{tabular}{@{}lrrrr@{}}
\toprule
Scale & Regions & Country codes & Eligible triplets & Mean $\Oinfo$\\
\midrule
ADM1 & 2,654 & 163 & 7,458,355 & 0.076012\\
NUTS-1 & 124 & 37 & 31,826 & 0.192506\\
NUTS-2 & 327 & 37 & 731,645 & 0.172919\\
NUTS-3 & 1,517 & 37 & 120,354,329 & 0.233174\\
\bottomrule
\end{tabular}
\end{table}

NUTS-3 uses the common European country-specific raster footprint. Its
mean TC/DTC are $1.294341/1.061167$ nats, with $10.6181\%$ negative
O-information; 42,860 geographic triplets fail numerical eligibility.
Median nearest-neighbour representative-point spacing is 34.302~km.
No NUTS-3 random reference is estimated. City occupancy and regional raster
aggregation are distinct: a region containing one sampled city need not
have that city's growth series. Global ADM2 is outside scope because
administrative levels do not identify a common scale or function worldwide.

\subsubsection{Generating matched random subdivisions}

At each scale, each country receives exactly its observed number of seeds;
Voronoi cells are constructed in Mollweide coordinates and clipped to its
retained footprint. Candidate sites are population-raster blocks of
1--10~km, represented by their highest-population pixel. The first seed is
drawn proportional to population; subsequent draws weight population by
distance to the nearest selected seed raised to $\beta$. This favours
occupied areas while separating seeds. The exponents are $0.5$, $1$ and
$0.5$ for ADM1, NUTS-1 and NUTS-2, selected as the first passing value in
geography-only calibration over $\{0,0.5,1,2,4\}$.

Candidates require positive GDP coverage in every year, without selecting
on GDP magnitude, growth or information. Each map must preserve country
region counts and native pixel ownership, with median nearest-neighbour
spacing within 10\%, geographic triplet counts within 25\%, and one-,
two- and three-country triplet fractions within five percentage points of
observed values. Spacing uses representative points; triplets use centroids.
Gates apply to both full and country-restricted geography before information
estimation. Every region needs positive finite population and GDP and
nonconstant growth; country population and annual GDP totals must agree
with the observed map to relative tolerance $10^{-10}$. Invalid maps are
rejected without reducing coverage. The archive documents occupancy
handling, retries, seeds and geometry repairs.

All 600 retained maps pass. The design was refined after earlier descriptive
maps and is not a prospective preregistration; geometry and acceptance rules
were fixed before production information estimates. Calibration and pilot
maps are excluded from the 200 maps per scale.

\subsubsection{Weighting and matched results}

Each map mean weights eligible triplets equally; each random map then
receives equal weight. Central 95\% ranges use linearly interpolated
2.5th/97.5th percentiles of map means. The statistic $z$ divides observed
minus random mean by the sample standard deviation across maps. The
lower-tail comparison is $(1+\#\{\overline O_{random}\leq
\overline O_{observed}\})/201$, without tail extrapolation.

\begin{table}[tbp]
\centering\small
\caption{Observed O-information relative to 200 matched maps per scale,
in nats. Standardised rows weight one-, two- and three-country triplet
means by their proportions in the observed map. Ranges describe random
maps, not sampling confidence intervals.}
\label{tab:partition-null}
\begin{tabular}{@{}llrrrr@{}}
\toprule
Scale & Weighting & Observed & Random mean & 95\% range & $z$\\
\midrule
ADM1 & Triplets & 0.076012 & 0.076442 & 0.074286--0.078695 & -0.345\\
ADM1 & Composition & 0.076012 & 0.075990 & 0.074160--0.077751 & +0.023\\
NUTS-1 & Triplets & 0.192506 & 0.196653 & 0.188377--0.203297 & -1.028\\
NUTS-1 & Composition & 0.192506 & 0.197655 & 0.188990--0.204888 & -1.283\\
NUTS-2 & Triplets & 0.172919 & 0.172289 & 0.165398--0.177347 & +0.206\\
NUTS-2 & Composition & 0.172919 & 0.172004 & 0.165337--0.177337 & +0.282\\
\bottomrule
\end{tabular}
\end{table}

Six samples---full, country-only, numerical-only, combined primary, and
combined with eigenvalue cutoffs $10^{-8}$ or $10^{-4}$---are compared
with raw triplet weights and weights standardised to observed country
composition. All 36 O-information comparisons lie inside their central
95\% random-map ranges; maximum absolute $z$ is 1.2834. Primary raw
lower-tail comparisons are 0.368, 0.144 and 0.577. This tests subdivision
within fixed countries, rather than causal international-border effects.

For ADM1, observed TC/DTC are $0.745541/0.669529$ nats versus random means
$0.746216/0.669774$. NUTS-1 gives $1.188811/0.996305$ versus
$1.222219/1.025566$, and NUTS-2 $1.098524/0.925604$ versus
$1.102274/0.929986$. NUTS-1 TC/DTC lie relatively low: lower-tail values
are 0.045/0.040, or 0.030/0.025 after composition standardisation. These
secondary unadjusted comparisons are not multiplicity-adjusted discoveries;
similar O-information does not imply identical positions for every measure.

\subsubsection{Whole countries versus one random map}

\begin{figure}[htbp]
\centering
\includegraphics[width=\textwidth,height=.58\textheight,keepaspectratio]{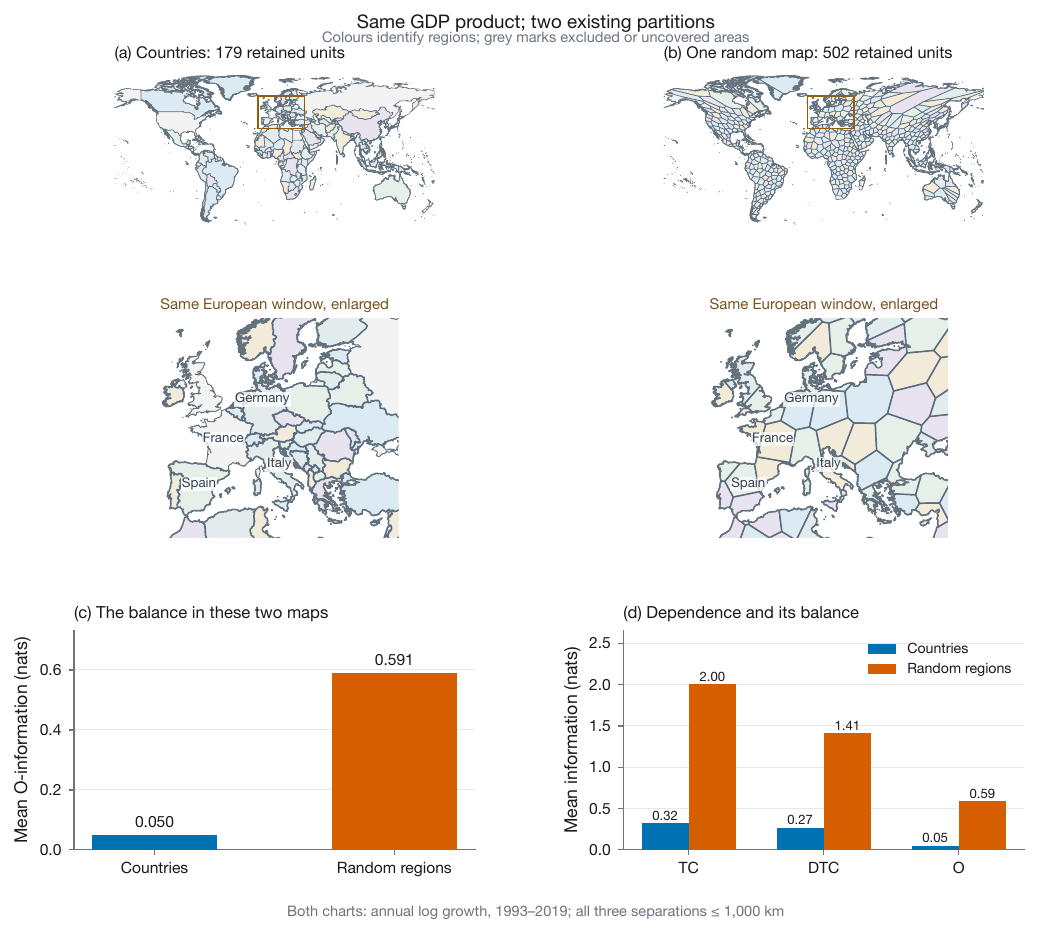}
\caption{Where the lines are drawn changes the measured growth relationships.
  Panels (a,b) show the country and random Voronoi partitions
  used in both charts, with the same European window enlarged. Colours
  identify regions; grey areas have no retained series or are uncovered.
  Regions require positive GDP per capita in all 28 years. This leaves
  179 countries and 502 random regions, contributing 1,354 and 2,963 triplets,
  respectively. Panels (c,d) use annual log growth for 1993--2019, the same
  Gaussian-copula estimator, and a 1,000~km limit on all three centroid
  separations. Panel (c) compares mean O-information; panel (d) shows mean
  TC, DTC and their difference, O-information. TC measures overall dependence;
  TC and DTC do not separately measure redundancy and synergy. There is
  one random realization, with different coverage and region counts from
  the country partition. The figure is an illustration, not a matched
  randomization test.}
\label{fig:partition-illustration}
\end{figure}

Figure~\ref{fig:partition-illustration} serves a different purpose. It
compares 179 countries with 502 random regions that can cross borders,
without matched coverage/counts or the primary country/numerical screen.
Median nearest-neighbour spacing is 430 versus 409~km. Similar spacing
does not equalise composition or coverage, and one random map supplies
no ensemble probability. The country geometry filter omits the United
Kingdom, France, Russia, United States, Antarctica, Fiji, New Zealand and
Kiribati for longitude spans over $180^\circ$, and Egypt for invalid
geometry; further units fail population/GDP eligibility. The illustration
is not a census of countries.

The harmonised primary country sample has 163 units and 1,305 triplets,
with mean O-information $0.05125$ nats. Excluding countries below
1,000~km$^2$ leaves 150 units/894 triplets and mean $0.04277$; requiring
one million attributed residents leaves 134/695 and mean $0.04221$.
No country-level matched null is reported: one cell per fixed country
would reproduce that country's boundary exactly.

\FloatBarrier

\subsection{Exploratory economic associations}
\label{app:delta-blocks}
\label{app:delta-convergence-table}
\label{app:delta-nested}
\label{subsec:delta-variance}
\label{app:delta-did-noise}
\label{app:delta-did-dyadic}

\subsubsection{Growth, participation and regional controls}

The convergence outcome is $\log GDPpc_{2019}-\log GDPpc_{2010}$.
Controls are 2010 log GDP per capita, industry/services/agriculture
employment-hours shares (construction is omitted), tertiary attainment
among ages 25--64 and the European Quality of Government Index (EQI).
Sources are Eurostat \texttt{nama\_10r\_2emhrw} and
\texttt{edat\_lfse\_04}, and the 2010 EQI wave in the 2024 time-series
release \citep{charron2014}. City points match NUTS-2024 regions for
Eurostat and NUTS-2021 regions at EQI's reported level 0, 1 or 2.
Missing matches stay missing; national EQI scores are used only where
the source itself reports them. Retrospective EQI harmonisation can copy
parent-region scores into later subregions. Complete controls retain
465 of 492 EU-27 cities, in 25 countries; Ireland and Lithuania lack
required regional education data.

Participation is the zero-to-one fraction of eligible incident triplets
with negative O-information under the primary restrictions. These growth
observations overlap the regression outcome. For city $i$ with incident
triplets $\mathcal T_i$, an additional control is
$a_i=\left|\frac{1}{|\mathcal T_i|}\sum_{t\in\mathcal T_i}\Oinfo_t\right|$, the absolute
mean score, not the mean absolute score. Incident triplets need not lie
wholly within the regression sample. The decile restriction retains
cities strictly above the complete-case tenth percentile of $a_i$
($0.0479838$ nats), leaving 418 cities. Added $R^2$ compares models
with identical rows and controls, omitting participation alone.
In Table~\ref{tab:delta-convergence}, HC1 denotes heteroskedasticity-robust
standard errors and ML denotes maximum-likelihood estimation.

\begin{table}[tbp]
\centering
\caption{Participation associations with 2010--2019 growth.
Coefficients are per unit of the zero-to-one participation fraction, not
standardised coefficients. Country-cluster rows use country-based degrees of
freedom; the absolute-O and decile rows use country clustering. Added $R^2$
compares nested OLS models on identical complete cases. Spatial-error
uncertainty uses the Gaussian likelihood. All results are associations.}
\label{tab:delta-convergence}
\begin{tabular}{lrrrrr}
\toprule
Specification & $n$ & Coef. & SE & $p$ & Added $R^2$ \\
\midrule
No lag; HC1 & 465 & -0.00113 & 0.06165 & 0.985 & 0.000001 \\
No lag; country clusters & 465 & -0.00113 & 0.13361 & 0.993 & 0.000001 \\
Spatial-error ML & 465 & +0.02362 & 0.03926 & 0.547 & --- \\
Add absolute mean O & 465 & -0.09894 & 0.10091 & 0.337 & 0.006033 \\
Drop bottom absolute-O decile & 418 & +0.07933 & 0.10295 & 0.450 & 0.004502 \\
\bottomrule
\end{tabular}
\end{table}

The spatial-error model is $y=X\beta+u$, $u=\lambda Wu+\epsilon$.
$W$ uses inverse great-circle distances to all other complete-case cities,
zero diagonal and rows summing to one. There is no distance cutoff or
spatial-lag-of-growth regressor. Estimated $\lambda=0.988$; its uncertainty
assumes this Gaussian spatial-error specification. Removing the bottom
decile excludes cities, not edges, and does not identify noisy observations.
Full-reference participation gives the same substantive conclusion.

\subsubsection{Four short PID windows}

The exploratory comparisons use GDP levels for 1992--1998, 1999--2005,
2006--2012 and 2013--2019: six within-window log changes each, excluding
changes across windows. Country/numerical restrictions retain 144,478
contexts in all four windows, aggregated to 75,731 pairs and 302,924
pair-block observations. Longitudinal estimates use this fixed balanced
sample. Selecting on all windows can depend on post-entry outcomes.
Many city pairs do not increase the observations in any one decomposition;
numerical consistency cannot establish reliability with six observations.

For each pair and window, information components are averaged over its
retained contexts. Synergy share divides mean synergy by mean joint target
information, rather than averaging context-level ratios or counting
component dominance. The primary gravity regression relates this share to
log distance and the sum of endpoint log total GDP in 1992. Coefficients
are $+0.006563$ (SE $0.006149$) and $-0.004504$ (SE $0.001921$), with
$R^2=0.00833$. Block indicators leave these slopes unchanged; the per-capita
variant gives $+0.007361$ and $-0.005827$. Uncertainty allows observations
sharing an endpoint or context country to be related. These small
associations do not identify a gravity mechanism.

\subsubsection{Institutional accession}

Following the cohort-and-time structure of \citet{callawaysantanna2021},
each cohort comprises pairs first jointly belonging to an institution in
the same block. We compare their changes with not-yet/never-joint-member
pairs, relative to the block before entry. A linear regression of outcome
changes on log distance in the control pairs supplies adjustment.
Supported cohort-time comparisons are weighted by treated-pair counts.
This structure does not establish causal identifying assumptions here.

Entry is assigned by membership at the block's final year. The primary
post-entry average omits that whole block; an alternative includes it.
Always-treated pairs lack a pre-entry reference, and the final cohort has
no later block for the entry-omitted average. Membership dates follow the
EU/ECB and European Commission chronologies; Schengen uses operational
border-control abolition, including Greece in 2000.\footnote{Sources:
\url{https://www.ecb.europa.eu/euro/intro/html/index.en.html} and
\url{https://home-affairs.ec.europa.eu/policies/schengen/schengen-area/history-schengen_en}.}
The restricted control pool requires both endpoints, but not the context,
in the 32-country membership-data list, leaving 12,149 balanced pairs.
The archive lists countries and cohort-specific support. This source list
is not the number of effective controls, which varies by cohort and time.

\begin{table}[tbp]
\centering
\caption{Descriptive adjusted changes in synergy share, in percentage points.
Global controls are not-yet/never-treated pairs in the balanced panel; the
restricted pool requires both endpoints in the 32-country treaty source.
The columns change entry-block handling and the control pool. They are descriptive adjusted comparisons.}
\label{tab:delta-did}
\begin{tabular}{lrrr}
\toprule
 & \multicolumn{2}{c}{Global controls} & Treaty-country controls \\
Membership & Entry included & Entry omitted & Entry omitted \\
\midrule
EU & -0.67 & +1.65 & -0.86 \\
Schengen & +6.36 & +13.53 & +28.59 \\
Euro area & +1.31 & +6.63 & +4.68 \\
\bottomrule
\end{tabular}
\end{table}

\begin{figure}[htbp]
\centering
\includegraphics[width=\textwidth,height=.58\textheight,keepaspectratio]{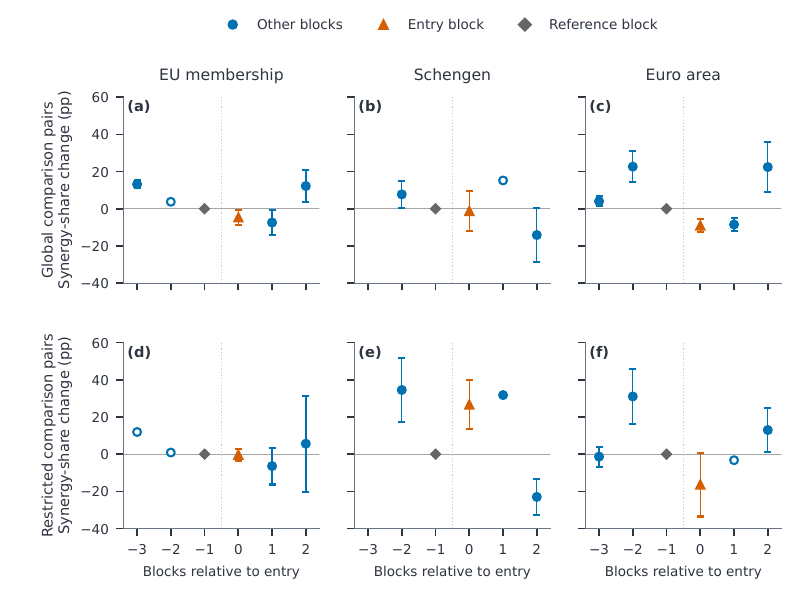}
\caption{\textbf{Accession comparisons depend on the choice of comparison countries.}
This exploratory diagnostic asks whether synergy's share of information
about a target city's growth changes when both countries in a cross-border
pair become members of the EU, Schengen area or euro area (columns).
The share is mean synergy divided by mean joint target information across
the pair's retained comparison cities.
Points show regression-adjusted changes relative to the preceding block,
compared with changes for pairs whose countries are not yet, or never,
joint members. Values are in percentage points. The top row uses comparison
pairs worldwide; the bottom row restricts both endpoint countries to the
32-country membership-data sample.
The horizontal axis counts seven-year blocks. Block 0 is the first classified
as jointly belonging to the institution by its final year; it can contain
both pre-entry and post-entry years (orange triangles). The preceding block
($-1$, grey diamond) is the zero reference. Blue circles show other blocks.
Error bars are the estimate plus or minus 1.96 standard errors, allowing
observations sharing countries to be related. Hollow
points lack a usable uncertainty estimate.
\textbf{Each PID estimate uses only six annual growth observations.}
Numerical checks do not establish that such short-series estimates are
statistically reliable. The changing patterns across comparison groups,
limited pre-entry evidence and short series provide no robust evidence of
a common membership effect.}
\label{fig:accession-diagnostic}
\end{figure}

Uncertainty uses contributions to the adjusted estimate, including the
estimated control regression. Covariance counts overlapping endpoint/context
country sets once and assumes independence between disjoint sets.
Figure~\ref{fig:accession-diagnostic} shows estimates plus or minus 1.96
standard errors; hollow points lack usable variance estimates.
The global entry-omitted Schengen variance is negative, so no interval is
reported or clipped to zero. For the global entry-omitted averages,
the EU and euro-area model intervals span zero. Deleting countries gives
ranges for these global entry-omitted averages of $-8.64$ to $+3.73$
percentage points for EU,
$+4.56$ to $+16.66$ for Schengen and $-1.94$ to $+10.34$ for euro
membership. These are sensitivity ranges, not confidence intervals.

Few joining countries limit uncertainty assessment. The event-$-3$ EU
comparison involves Croatia alone as a joining country; the euro comparison
involves Latvia and Lithuania. The first cohort lacks another pre-period
before its reference block. Such limited pre-entry evidence cannot
establish parallel trends. Combined with short-series PID and control-pool
sensitivity, the diagnostic provides no robust common membership effect.

\FloatBarrier

\subsection{Code and supporting data}
\label{app:repro}

The analysis code and derived results are available from the corresponding
author upon reasonable request. A public reproducibility archive will be
released upon journal publication. Its results index links figures and
tables to inputs, samples, saved outputs and reproduction instructions.
Records preserve city identities, context selections, partition assignments,
exclusions and regional-control crosswalks. A validation-notes index locates
precision checks, calibration, sample-length diagnostics, geometry handling
and model verification, with omitted auxiliary tables and sensitivities.
Archive documentation records availability and provider attribution.

\FloatBarrier

\FloatBarrier
\end{document}